\documentclass[aps,prb,twocolumn,superscriptaddress,groupedaddress]{revtex4} 

\usepackage{graphicx}  
\usepackage{dcolumn}   
\usepackage{bm}        
\usepackage{amssymb}   
\usepackage{float}
\usepackage{booktabs,array}

\usepackage{amsmath,mathrsfs}
\usepackage{amsthm}
\usepackage{epsfig}

\usepackage{color}
\usepackage{soul}
\usepackage{ulem}

\begin{document}

\title{Band structure engineering of ideal fractional Chern insulators}

\author{Ching Hua Lee}
\affiliation{Institute of High Performance Computing, 138632, Singapore}
\email{calvin-lee@ihpc.a-star.edu.sg}
\author{Martin Claassen}
\affiliation{Department of Applied Physics, Stanford University, Stanford, CA 94305, USA}
\author{Ronny Thomale}
\affiliation{Institute for Theoretical Physics and Astrophysics, University of W\"urzburg, Am Hubland, D-97074, Germany}

\date{\today}

\begin{abstract}
As lattice analogs of fractional quantum Hall systems, fractional Chern insulators (FCIs) exhibit enigmatic physical properties resulting from the intricate interplay between single-body and many-body physics. In particular, the design of ideal Chern band structures as hosts for FCIs necessitates the joint consideration of energy, topology, and quantum geometry of the Chern band. We devise an analytical optimization scheme that generates prototypical FCI models satisfying the criteria of band flatness, homogeneous Berry curvature, and isotropic quantum geometry. This is accomplished by adopting a holomorphic coordinate representation of the Bloch states spanning the basis of the Chern band. The resultant FCI models not only exhibit extensive tunability despite having only few adjustable parameters, but are also amenable to analytically controlled truncation schemes to accommodate any desired constraint on the maximum hopping range or density-density interaction terms. Together, our approach provides a starting point for engineering ideal FCI models that are robust in the face of specifications imposed by analytical, numerical, or experimental implementation. 
\end{abstract}

\maketitle

\section{Introduction} Lattice realizations of the fractional quantum Hall effect (FQHE) date back to a time not much later than its discovery in two-dimensional electron gases at low carrier density and strong magnetic field~\cite{tsui-82prl1559,laughlin83prl1395}. In 1987, inspired by an idea of D.~H.~Lee, Kalmeyer and Laughlin found that the bosonic FQHE Laughlin state at magnetic filling $\nu=1/2$ can be interpreted in terms of spin flip operators acting on a half-filled spin-polarized reference vacuum, the first instance of the chiral spin liquid~\cite{kalmeyer-87prl2095} (CSL). 
Although the CSL was first described for a Mott state on a triangular lattice~\cite{kalmeyer-89prb11879}, it was soon observed that the CSL can materialize on {\it any} lattice arrangement of discrete points~\cite{zou-88prb11424}, since it is fully constrained by the holomorphicity inherited from the Laughlin polynomial and its universal spin singlet character~\footnote{A mathematically more rigorous proof of the latter was provided based on a generalised Perelomov's identity~\cite{PhysRevB.85.155145}.}. The CSL laid the foundation for the concept of topological order~\cite{wen-89prb7387}, and was intensely discussed in the context of anyon superconductivity~\cite{wen-89prb11413}. Only more than a decade latter it was found that exact parent Hamiltonians for the CSL do exist~\cite{schroeter-07prl097202,thomale-09prb104406}, and that they can be generalized to a series of SU(2)$_{k}$ non-Abelian chiral spin liquids~\cite{greiter-09prl207203,PhysRevB.89.165125,PhysRevB.91.241106}, in close analogy to the Read-Rezayi series of the FQHE~\cite{Read-99prb8084,PhysRevB.89.085101}. The analytic construction of Ref.~\onlinecite{schroeter-07prl097202} straightforwardly carries over to a long-range hopping lattice model for interacting bosons~\cite{PhysRevLett.105.215303}, which accordingly features the exact parent Hamiltonian property for the $\nu=1/2$ Laughlin state. 

Several new facets to the field of lattice FQHE have since been discovered with the advent of fractional Chern insulators (FCIs)~\cite{tang-11prl236802,neupert-11prl236804,sun-11prl236803,PARAMESWARAN2013816,rev-emil} in 2011. FCIs are analogs of FQH states putatively realized in fractionally filled bands of lattices without orbital magnetic field~\cite{Venderbos2011,Hu2011,Parameswaran2012,Sheng2011,wang2011fractional,Wu2012,Goerbig2012,Roy2012,Murthy2012,Neupert2011a,Neupert2012,Venderbos2012,Kourtis2012,Wu2012a,Jain1989,Liu2013,lauchli2013hierarchy}. In particular, Abelian and non-Abelian FCIs can be realized not just for bosons but also for fermions, and, as opposed to the Abelian and non-Abelian CSLs, are not constrained to half-integer lattice fillings. While any substantiated hope for the realization of a CSL has so far been limited to frustrated magnetisation plateaus~\cite{PhysRevB.90.174409} or optical lattice setups~\cite{aebn,normmike}, the FCIs promise a broad range of systems where they could potentially be realized, such as in strongly correlated multi-orbital Hubbard models~\cite{PhysRevLett.108.126405}, transition metal oxides\cite{xiao2011interface}, cold atoms lattices\cite{cooper2013reaching,yao2013realizing,maghrebi2015fractional}, and dynamically driven systems\cite{hafezi2013non,grushin2014floquet}.  Furthermore, while any single-particle features are suppressed in the CSL construction by assuming a Mott insulator from the outset, the band structure is of outmost importance in FCIs. In fact, the single-particle parameters turn out to be as relevant as the form of interactions in giving rise to topologically ordered many-body ground states in Chern bands. 

As a new set of model parameters, the band structure in FCIs transcends the realm of conventional FQHE. Firstly, with the lattice scale $a$ replacing the magnetic length $l_{\text{B}}$ from FQHE, FCIs are, in principle, expected to exhibit large bulk energy gaps and be observable at higher temperatures. Secondly, the source of Berry flux does not derive from an external magnetic field, but from the Chern band itself. On the Chern lattice, various additional factors such as inter-band mixing and multi-orbital composition contribute to determining the effective Berry flux profile of the Chern band, as well as giving rise to a much richer landscape of band interference mechanisms than from Landau level mixing in FQHE. Finally, FCIs on bands with Chern number $C>1$ exhibit a myriad of Abelian and non-Abelian phases beyond those from the lowest Landau level. They promise higher accessibility and tunability than their multi-layer FQHE counterparts, which they map to under suitable limits~\cite{PhysRevX.2.031013} and accurate matching of boundary conditions~\cite{PhysRevLett.110.106802,lee2014lattice}.  

The band structure of FCIs, however, also poses further challenges in the engineering of optimized models that stabilize topologically ordered quantum states of matter.  So far, the central hypothesis has been that the type of topological order we expect to find in FCIs should at least be guided, if not fully constrained, by (multi-layer) FQHE. As a consequence, the predominant task has come to be the design of optimal conditions for fractional quantum Hall fluids to exist in analogous FCI lattice models. If not for the large typical values of $l_{\text{B}}$,
favorable properties such as the homogeneous Berry flux by the external field, the rotation invariance of the electron gas continuum, the high mobility (i.e. low carrier density and low disorder) experienced by the electrons, and the perfect flatness of Landau levels would all render the FQHE an ideal setup for such fluids. This is why we desire to replicate as much as possible of these properties in an FCI lattice environment. 

Following the guiding principle outlined above, the main difficulty lies in optimizing all Chern band parameters simultaneously in one physically sensible lattice model. In earlier attempts, for instance, it has been realized that perfectly flat quasi-Landau level bands can be written down by employing the previously developed machinery for CSL parent Hamiltonians~\cite{schroeter-07prl097202,PhysRevLett.105.215303}. These can be applied in FCIs to derive exact parent Hamiltonians taking identical forms as those in second-quantized FQHE models (see e.g. Refs.~\onlinecite{lee-04prl096401,seidel-05prl266405,PhysRevX.5.041003,yang2016generalized}). A main deficiency in this attempt, however, is that the band structure violates locality due to long-range hoppings. It does not help us in identifying optimal local FCI hopping models amenable to experimental realization. Rather, one needs to carefully balance considerations of band energetics, topology, and geometry, namely regarding the uniformity of band dispersion, Berry curvature, and quantum distance, respectively. Bringing these aspects together in an optimized band structure engineering scheme, which can also prove useful as a guiding principle for experimental band structure design, is the central focus of our investigation.

In this article, our starting point will rest on two of our previous works, one on the systematic treatment of band flattening through the imaginary gap of the given Chern band structure (Ref.~\onlinecite{lee2016band}), and the other on introducing momentum-space guiding-center geometry as a new variational degree of freedom for preserving locality of pseudopotentials in FCIs with discrete rotation symmetry (Ref.~\onlinecite{claassen2015}). 
We combine these sources of insight to engineer a class of ``ideal'' FCI lattice models hosting Chern bands that act as analogs of the isotropic Landau level, while displaying optimally flat band dispersion and uniform Berry curvature $F_{xy}$. In addition, we optimize the quantum band geometry of the problem. In Ref.~\onlinecite{claassen2015}, we have shown that an FCI lattice obeying the ``Ideal Droplet Condition'' $F_{xy} =2\sqrt{\text{Det}\, g}$, where $g$ denotes the guiding-center (or Fubini-Study) metric, can be described in an optimally-localized guiding-center basis that recreates an FCI analog of an anisotropic Landau level in a spatial magnetic field. The effective dynamics are captured by a first-quantized Hamiltonian that is the dual of the conventional FQH scenario, with the roles of position and momentum interchanged. In the following, we will predominantly concern ourselves with the subclass of isotropic FQHE analogs, where discrete rotation symmetry promotes the ``Ideal Droplet Condition'' to the stronger ``Ideal isotropic FCI condition'' $F_{xy} = \textrm{Tr}\, g$. If this condition is met, Haldane pseudopotentials decompose into local and nearest-neighbor interactions on the FCI model, yielding an excellent lattice representation of any desired isotropic FQHE model\footnote{But see Refs. \onlinecite{yang2012band}, \onlinecite{yang2012model} and \onlinecite{yang2016generalized} for an in-depth treatment of anisotropy in the FQHE.}. All these findings illustrate that band dispersion, topology, and quantum geometry have to be arranged in a concerted fashion to accomplish ideal lattice analogs of the isotropic FQHE~\footnote{From a complementary perspective, this likewise becomes beautifully transparent at the level of non-commutative projected FCI density operators~\cite{PhysRevB.85.241308,PhysRevB.90.165139}.}.

The strategy of our paper is as follows: (1) We show that the ``Ideal isotropic FCI condition'' $F_{xy} = \textrm{Tr}\,g$ is satisfied by a unique class of lattice models with Bloch states built from doubly-periodic meromorphic (elliptic) functions whose $C$ poles can be interpreted as momentum-space ``instantons'' corresponding to Berry curvature charge. Such models are fully determined by these poles, each characterized by only a small number of adjustable attributes. (2) We demonstrate that strong constraints on the few model variational parameters allow for efficient minimization of the non-uniformity of the Berry curvature and hence $\textrm{Tr}\,g$. Together, they lead to an improved preservation of magnetic translation symmetry, and as such a better agreement of the FCI density algebra with the FQH scenario up to third order\cite{Roy2012}. (3) We show that, due to the holomorphicity of our Bloch wavefunctions\cite{jian2013momentum}, the `Ideal isotropic FCI condition'' as well as the uniformity of $F_{xy}=\textrm{Tr}\,g$ are still mostly preserved once the resulting hopping Hamiltonian is real-space truncated into a physically realistic short-ranged lattice model. This gives our approach a decisive edge to connect ideal theoretical descriptions and experimentally relevant Chern band models.

The paper is organized as follows. In Section II, to present our approach in a self-contained form, we introduce the Chern band description and our geometric guiding-center description of FCIs in a detailed and most accessible fashion. In particular, we rederive the "Ideal isotropic FCI condition'' from first principles, which is the starting point for us to specialize to holomorphic guiding-center metrics in Section III. Whenever suitable, technical details are delegated to the appendices. In Section IV, the explicit optimization scheme is detailed in a step-by-step fashion, including a treatment of the optimized truncation of the hopping range through complex analytic properties of the Bloch functions. For two-band models, the Ideal isotropic FCI condition takes on an elegant geometric form on the Bloch sphere. This conceptual discussion is complemented by exemplary explicit models presented in Section V, where we demonstrate how our optimization scheme is implemented. As found therein, our optimization scheme yields excellent band structure setups for Chern bands of arbitrary $C>1$ and number of bands $N$, where physically realistic Hamiltonians emerge after appropriate real-space truncation. One of the central accomplishments manifests in a three-band, $C=3$ Chern model with almost uniform band dispersion {\it and} Berry curvature satisfying $\text{Tr}\,g=F_{xy}$, as well as further models possessing particular uniformity in energy, Berry curvature, and quantum geometry. In Section VI, we conclude that by enforcing the desired quantum band geometry from the outset, we have managed to define a promising scheme that optimizes band flatness and Berry curvature homogeneity on the same footing. As such, it produces prototypical ideal Chern band structures for the stabilization of FQH fluid analogs in FCIs.

\section{Preliminaries}

\subsection{Chern Insulators and Band Topology}

Generically, the Hamiltonian for a fractional Chern insulator (FCI) takes the form
\begin{eqnarray}
H&=&H_{0}+ H_{\text{int}}\label{FCI},\notag\\
H_0&=& \sum_{\substack{k\in \text{BZ} \\ \alpha\beta} 
}H_{\alpha\beta}(k)c^\dagger_{\alpha k }c_{\beta k}^{\phantom{\dagger}},
\end{eqnarray}
where $\alpha,\beta=1,...,N$ are spin/orbital indices of $H_{\alpha\beta}(k)$, the matrix elements for the $N\times N$ single-body Hamiltonian matrix $H(k)$ for $N$ bands. $H_{\text{int}}$ represents the many-body interaction which is the analog of the Coulomb interaction in FQH systems. In this paper, we shall focus on engineering a ``ideal'' lattice analog of the FQH at the bandstructure level, i.e. through the single-body Hamiltonian $H(k)$. With this choice of the ``ideal'' $H(k)$, locality in the interaction $H_{\text{int}}$ directly carries over to the FQH pseudopotentials, and are chosen according to the desired FQH state to be realized\cite{lee2013,lee2014lattice,PhysRevX.5.041003,yang2016generalized}. 

Let $\epsilon_n$ and $|\varphi_n\rangle$ denote the eigenenergy and the (periodic part of the) normalized Bloch eigenstate of the $n^{th}$ band of $H(k)$: $H(k)|\varphi_n(k)\rangle = \epsilon_n(k)|\varphi_n(k)\rangle$, $n=1,2,...,N$. The simplest route to fractional Chern insulator physics involves a single fractionally-filled isolated band $n$ that is well-separated from filled bands of lower energy and unoccupied bands of higher energy. This band $n$ is characterized by the two geometric quantities $	F^{(n)}_{xy}$ and $	g^{(n)}_{ij}$ given by\footnote{Incidentally, we shall also use $Q$ as the starting point of our FCI band Hamiltonian.}
\begin{equation}
g_{ij}^{(n)}+iF^{(n)}_{ij}= \langle\partial_i\varphi_n|Q|\partial_j \varphi_n\rangle,
\label{Q}
\end{equation}
$Q=\mathbb{I}-|\varphi_n\rangle\langle\varphi_n|$, which can be expressed in the more familiar form 
\begin{equation}
	F^{(n)}_{xy} = \partial_{k_x} A^{(n)}_y - \partial_{k_y} A^{(n)}_x ,\end{equation}
where $A^{(n)}_j = -i\langle \varphi_n|\partial_j\varphi_n\rangle$, $j=k_x,k_y$, and
\begin{equation}
	g^{(n)}_{ij} = {\rm Re}\left\{ \langle\partial_i\varphi_n|\partial_j \varphi_n\rangle- \langle \partial_i\varphi_n| \partial_j\varphi_n\rangle \right\}.
\label{FG}
\end{equation}
Intuitively, the Berry curvature $F_{xy}^{(n)}$ is  proportional to the phase accrued as $|\varphi_n\rangle$ is brought around a small loop in the Brillouin zone (BZ); $F_{xy}^{(n)}$ is thus a measure of the non-commutativity of the band geometry. 

The less commonly known $g_{ij}^{(n)}$, which is known as the Fubini-Study metric, characterizes the {quantum distance} $D(k,k+dk)$  between two points $k$ and $k+dk$ in the BZ. Intuitively, it is a measure of how much the band "changes" across these two points\cite{marzari1997maximally,lee2014lattice}. A basis-invariant way to characterize a band is via the projector $P_k=|\varphi_n(k)\rangle\langle \varphi_n(k)|$ onto its eigenstate at a point $k\in \text{BZ}$. Hence the quantum distance between points $k$ and $k+\Delta k$ can in general be defined as the departure of $P_kP_{k+\Delta k}$ from the identity: 
\begin{equation}
D(k,k+dk)=\text{tr}[\mathbb{I}-P_kP_{k+dk}]=g_{ij}dk_idk_j,
\end{equation}
where $\text{tr}$ is the trace taken over the space of occupied eigenstates (bands $N_{\text{f}}$). In our case with just a single band \cite{marzari1997}, $D(k,k+dk)|_{N_{\text{f}}=1}=1-|\langle \varphi(k)|\varphi(k+dk)\rangle|^2$ and we recover the Fubini-Study metric defined in Eq. \ref{FG} above. Its geometric meaning will be further elaborated on in Subsection \ref{N2bands}; interestingly, it is also in principle physically manifested in the current noise spectrum\cite{neupert2013measuring}.

In the following, we will drop band indices $n$ and focus solely on the partially-filled isolated band. The topology at the level of band theory is characterized by the integer $C$ called Chern number
\begin{equation}
C=\frac1{2\pi}\int F_{xy} d^2k,
\end{equation}
which is nonzero for a nontrivial Chern Insulator. The Chern number is the winding number of the map from the BZ, which is a 2-torus $T^2$, to the complex projective plane~\footnote{In the case of $N_F$ occupied bands, it will be a map from the torus $T^2$ to the complex Grassmannian $\frac{U(N)}{U(N_F)\times U(N-N_F)}$. } $\frac{U(N)}{U(1)\times U(N-1)}$, where the set of eigenstates resides~\cite{ryu2010}. This is most easily visualized in the case of a 2-band model $H(k)=\vec d(k)\cdot \vec \sigma$ with one occupied band, where $\vec \sigma=(\sigma_1,\sigma_2,\sigma_3)$ is the vector of Pauli matrices. It maps $T^2$ to $\frac{U(2)}{U(1)\times U(1)}\sim S^2$ with Berry curvature $F_{xy}=\frac1{2}\vec d\cdot(\partial_x\vec d\times \partial_y \vec d)$. This is just the area on the Bloch sphere swept out by an unit area on the BZ. 

As such, $ F_{xy}$ can be interpreted as the ``Jacobian'' of this map, whose uniformity is quantified by the mean-square deviation
\begin{equation}
\langle(\Delta F)^2\rangle= \frac{1}{4\pi^2}\int \left( F_{xy} - \frac{C}{2\pi}\right)^2d^2k.
\label{dFxy}
\end{equation}
Note that $\langle(\Delta F)^2\rangle$ has a finitely large lower bound for $N=2$ bands, as it is impossible to have a map $T^2\rightarrow S^2$ that has a constant Jacobian, as can be easily seen by drawing a grid on these manifolds. However, $\langle(\Delta F)^2\rangle\rightarrow 0$ is theoretically achievable for $N\geq 3$, and we shall construct a specific example for it in Subsection \ref{N3}. 

\subsection{Position-Momentum Duality and the Geometrical Description of Fractional Chern Insulators}

\begin{figure}[h]
 \includegraphics[width=0.97\linewidth]{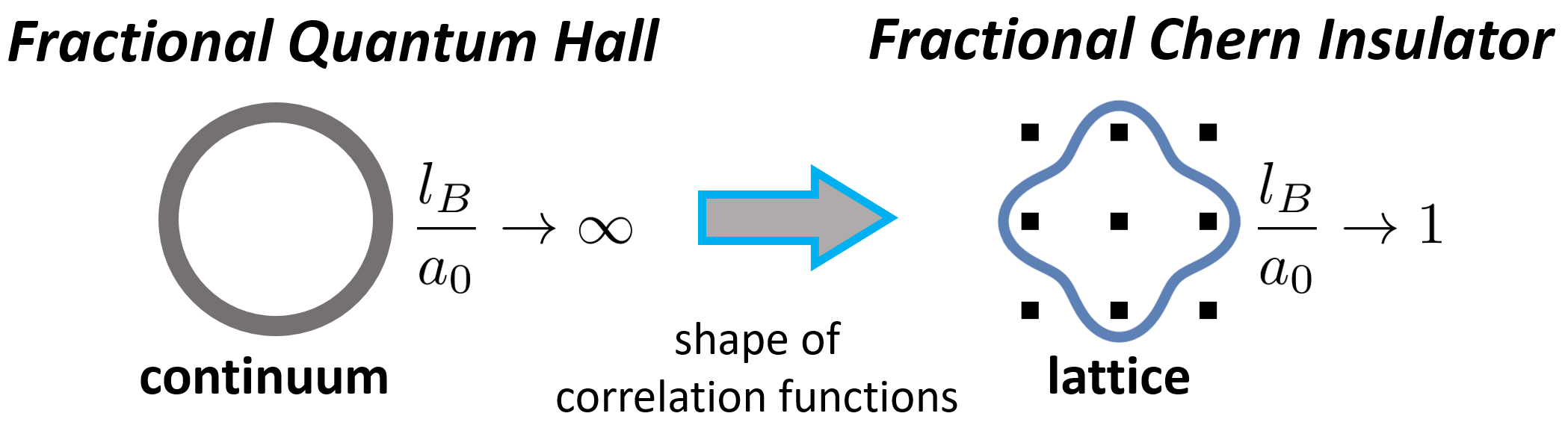}
\caption{(Color Online) Cartoon picture of the role of the lattice in affecting the geometry of a FQH fluid, which in turn determines the shape of correlation functions. The left figure depicts the rotationally-symmetric case of the conventional FQH effect, with magnetic length $l_B$ far exceeding the lattice constant $a_0$. The right figure depicts the opposite FCI limit, where the lattice constant is equal or commensurate with the magnetic length; as shown here, a $C_4$-symmetric distortion of the correlation functions should generically be expected upon placing the FQH fluid on the square lattice.
}
\label{fig:cartoon}
\end{figure}

We now review connections between geometric ingredients of the bandstructure and the geometry of FQH ground state wave functions in a flat Chern band. Motivated by analogies between a flat Chern band and the dispersionless Landau levels of the FQHE, it was found early on that the interacting problem on a fractionally-filled Chern band mirrors, at long wavelengths, the universal continuum limit of the FQH problem, being dictated solely by topology\cite{Parameswaran2012, Roy2012}. Numerical evidence, however, quickly accumulated for a variety of different lattice models, and indicated that two \textit{a priori} analogous Chern insulators with similar band flatness and distribution of $F_{xy}$ over the Brillouin zone might entail drastically different propensities to stabilize incompressible states at fractional filling. 

In our previous work~\cite{claassen2015}, it was shown that a crucial step to understanding this discrepancy is to consider instead the \textit{guiding center geometry} in the flat band, essentially discarding the notion of an \textit{isotropic} FQH fluid on the lattice. At its heart, this program traces back to seminal work by Haldane \cite{haldane2011} which demonstrated that isotropy is not essential to the FQHE. To connect these views, start from an isotropic free electron gas in a magnetic field, i.e., the canonical model of the FQHE:
\begin{align}
\hat{H} = \frac{1}{2m} \sum_i \left( \hat{\mathbf{p}}_i + e\mathbf{A} \right)^2 + \sum_{i<j} V(\hat{\mathbf{r}}_i - \hat{\mathbf{r}}_j)   .
\label{eq:HamMagneticField}
\end{align}
Here, $m$ is the effective mass, $\mathbf{A} = \frac{1}{2l_B} [-y, x]^\top$ denotes the vector potential, and we shall set the magnetic length $l_B = 1$. We constrain this discussion to filling fractions $\nu < 1$ and assume that the Landau level splitting (cyclotron frequency) of the single-body problem is much larger than the scale of interactions. In this case, energetics dictate that the many-body ground state must have vanishing probability of occupying the higher-lying Landau levels. In microscopic (but usually not in effective field theory) treatments, one thus canonically considers the projected dynamics within the lowest Landau level (LLL) only. Formally, the single-body problem can be subdivided into left-handed (Landau level) and right-handed (``guiding center") degrees of freedom. Introducing the corresponding ladder operators
\begin{align}
	&\hat{b}^\dag = ( -i \partial_{z} + i\bar{z}/2 )/\sqrt{2} \\
	&\hat{a}^\dag = (-i \partial_{\bar{z}} + iz/2 )/\sqrt{2}   \label{eq:isotropicGuidingCenterLadderOps}
\end{align}
in terms of complex coordinates $z = x+iy$, the single-body Hamiltonian reduces to $\hat{H}_0 = \omega_C (\hat{b}^\dag \hat{b} + 1/2)$, with the guiding-center $(\hat{a},\hat{a}^\dag)$ degeneracy apparent. In the LLL, corresponding wave functions for the single-body basis readily follow as $\Psi_m(z) \sim z^m e^{-|z|^2/4}$ with $m = 0,1,\dots$ indexing the guiding center. Crucially, they are simultaneous eigenstates of  (\ref{eq:HamMagneticField}) and canonical angular momentum, with $L_z = \hbar m$. (Note that the kinetic angular momentum is zero for all states in the LLL.) The familiar many-body trial wave functions can now be readily expressed in terms of guiding center operators; for instance, Laughlin's set of incompressible trial wave functions for filling fractions $\nu = 1/n$, with $n$ odd, read
\begin{align}
	\Psi_L^{\nu} = \prod_{i<j} \left( \hat{a}^\dag_i - \hat{a}^\dag_j \right)^{(1/\nu)} \Psi_0,
\end{align}
where $\Psi_0 = \exp\left( -\sum_i |z_i|^2 / 4 \right)$.

To proceed, it is crucial to observe that the choice of guiding-center basis $\Psi_m(z) \sim z^m e^{-|z|^2/4}$ is not unique. Instead, the degeneracy of the LLL affords, in principle, the construction of an \textit{arbitrary} guiding-center basis $\tilde{\Psi}_m(\mathbf{r})$ in the LLL, indexed by integers $m$, with the sole constraint for Laughlin states that the wave function vanish as a power of $1/\nu$ when two particles $i,j$ approach each other as $|r_i - r_j| \to 0$, with appropriate generalizations for more exotic FQH states. In the isotropic case, a choice of guiding center basis that respects continuous rotational symmetry seems obvious. However, already the simplest generalization -- electron gases with anisotropic in-plane effective mass -- breaks this picture. This can be implemented by promoting the mass in Eq. \ref{eq:HamMagneticField} to an effective mass tensor generically parametrized by a unimodular (``Galilean") metric\cite{haldane2011,yang2012band,yang2016generalized} $g^{\mu\nu}$:
\begin{align}
	\hat{H} = \frac{1}{2m} \sum_i \hat{\pi}_{i\mu} g^{\mu\nu} \hat{\pi}_{i\nu} + \sum_{i<j} V(\hat{\mathbf{r}}_i - \hat{\mathbf{r}}_j),
	\label{Hmetric}
\end{align}
where $\hat{\pi}_\mu = \hat{p}_\mu + e A_\mu$, $\mu = x,y$ are the gauge-invariant dynamical momenta. The spectrum is again described by Landau levels with macroscopic degeneracy. However, after abandoning isotropy, a natural generalization of the guiding-center operators is instead
\begin{align}
	&\hat{a}^\dag = \omega^\mu \left( -i \partial_\mu + i A_\mu \right),
	\label{hata}
\end{align}
where $\omega^\mu$ describes the deformation from the ``circle", and also defines a unimodular \textit{guiding-center metric}
\begin{align}
	\eta^{\mu\nu} = \omega^\mu \bar{\omega}^\nu + \bar{\omega}^\mu \omega^\nu
\end{align}
that describes the shape of correlation functions for the FQH state (Fig. \ref{fig:cartoon}). 
We assume that the Coulomb repulsion still retains continuous rotational symmetry $V(r) = V(|r|)$. If the electron gas is perfectly isotropic ($g^{\mu\nu} = \delta^{\mu\nu}$), the obvious choice for the guiding centers is the isotropic metric $\eta^{\mu\nu} = \delta^{\mu\nu}$, which corresponds to $\omega^x = 1/\sqrt{2},~ \omega^y = i/\sqrt{2}$, readily recovering Eq. (\ref{eq:isotropicGuidingCenterLadderOps}). If the electron gas is, however, anisotropic, the guiding-center metric $\eta^{\mu\nu}$ is pinned to neither the kinetic ($g^{\mu\nu}$) nor the interacting ($\delta^{\mu\nu}$) geometry. Instead, $\eta^{\mu\nu}$ will generically settle for a compromise between the two, and enters as a \textit{variational} degree of freedom that determines the geometry of the ground state wave function \cite{haldane2011,yang2012band,qiu2012}.

For a Chern insulator, in contrast to a free electron gas, the necessity to abandon isotropy is a quintessential consequence of placing a FQH fluid on the lattice. While a Bloch band indexed by momenta $k  \in \textrm{BZ}$ appears \textit{a priori} to have a structure significantly different to a LLL, an analogous guiding-center basis can nevertheless be constructed on the lattice. This is achieved by alternatively defining the FQH eigenbasis as eigenstates of an infinitesimal anisotropic confinement potential $\bar{V}$, projected to the flat Chern band \cite{claassen2015}. Explicitly, for FCIs, the projected confinement potential reads
\begin{align}
	\bar{V} = \tfrac{1}{2} \hat{\Pi}_\mu \eta^{\mu\nu} \hat{\Pi}_\nu + \tfrac{1}{2} \eta^{\mu\nu} g_{\mu\nu},
\end{align}
whose first term is of the same form as that of a FQH system (Eq.\ref{Hmetric}). Here $\eta^{\mu\nu}$, $\mu = x, y$ again plays the role of the guiding-center metric, and $g_{\mu\nu}$ is the Fubini-Study metric of the flat band. 
\begin{align}
	\hat{\Pi}_\mu = -i\partial_{k_\mu} + A_\mu(k )
\end{align}
are the \textit{momentum-space analogs} of the usual canonical momentum operators with $A_\mu$ denoting the Berry connection. The guiding-center basis now follows as eigenstates of $\bar{V}$.
To make this construction more transparent, we introduce again the metric decomposition $\eta^{\mu\nu} = \omega^\mu \bar{\omega}^\nu + \bar{\omega}^\mu \omega^\nu$, and define momentum-space ladder operators $\hat{\pi} = (-i\partial_{k_\mu} + A_\mu) \omega^\mu$, in analogy to the FQH guiding-center operators in Eq. \ref{hata}. The confinement operator then becomes
\begin{align}
	\bar{V} = \hat{\pi}\hat{\pi}^\dag + \frac{1}{2} \left( \eta^{\mu\nu} g_{\mu\nu} - \Omega \right) . \label{vpot}
\end{align}
By inspection, one can see that the first term in Eq.~\ref{vpot} just describes an electron in a magnetic field defined in momentum space, i.e. a region of nonzero Berry curvature. Its eigenstates are anisotropic \textit{momentum-space Landau levels} (MLLs) on the Brillouin zone torus, which in real space correspond to radially-localized wave functions on the lattice with Gaussian tails. Importantly, the roles of left-handed and right-handed degrees of freedom are interchanged: the MLL Landau level index now takes the role of the guiding-center index, whereas the $C$-fold degeneracy per MLL corresponds to an internal ``component" index for $C>1$ Chern bands.

Conversely, the right-most term in Eq.~\ref{vpot} is purely dispersive and spoils the radially-localized MLL basis by delocalizing the guiding-center wave functions on the lattice. The minimization of this term defines a \textit{preferred} guiding center metric $\tilde{\mathbf{\eta}}$ on the lattice given by
\begin{align}
	\tilde{\eta} = \sqrt{\det g} ~g^{-1},
\end{align}
which takes a role analogous to the Galilean metric in Haldane's formalism of the conventional FQHE. Among possible lattice models, there exists a set of optimal models for which (1) the dispersive contribution $\eta^{\mu\nu} g_{\mu\nu} - \Omega$ vanishes exactly, and (2) the guiding-center metric that is preferred by the lattice is isotropic ($\eta^{\mu\nu} = \delta^{\mu\nu}$). It was shown\cite{claassen2015} that these lattice models are exactly those that obey the  {ideal isotropic FCI condition}
\begin{align}
		F_{xy}(k ) = \textrm{Tr} \, g(k ). \label{eq:idealIsotropicDropletCondition}
\end{align}
This condition can be met by demanding that the Bloch functions $\left| u_k  \right>$ that span the flat Chern band are meromorphic functions $\frac{1}{N_k } \left| u_{k_x + i k_y} \right>$ in momentum, up to a normalization factor $N_k $. 
The search for such FCI models is thus of paramount importance, and the central focus of this paper is to study in detail how simple physical lattice models can provide excellent approximations of such ideal FCIs.

So far, the above discussion solely reviewed the construction of anisotropic guiding-center bases in the conventional FQHE and in FCIs, with no reference to the role of interactions. To connect the two, it was further shown that if the above ideal isotropic FCI condition (\ref{eq:idealIsotropicDropletCondition}) is met, the corresponding \textit{many-body problem} on the lattice has a particularly simple \textit{first-quantized} representation \cite{claassen2015}:
\begin{align}
	\hat{H}_{\text{int}} &= \sum_{i < j} V_{q} e^{i q_+ (\hat{\pi}_i - \hat{\pi}_j)} e^{i q_- (\hat{\pi}^\dag_i - \hat{\pi}^\dag_j)}.
\end{align}
Here, $q_{\pm} = q_x \pm i q_y$ in the isotropic case, and $V_{q} = \sum_l e^{iql} ~ V(l)$. Crucially, one can see by inspection that the many-body problem projected to the flat band entails an emergent conservation of center-of-mass of the guiding centers, with weak deviations due to Berry curvature fluctuations, averaged over the Brillouin zone:
\begin{align}
	\frac{A_{\textrm{BZ}}}{2} \left\| [ \hat{\Pi}_{\rm rel} , \hat{\Pi}^\dag_{\rm cm} ] \right\|^2 = \int_\textrm{BZ} d^2k ~ \left[ \Omega(k ) - \frac{2\pi C}{A_{\textrm{BZ}}} \right]^2 .\end{align}
Here, $\hat{\Pi}_{\rm cm/rel} = (\hat{\Pi}_1 \pm \hat{\Pi}_2)/\sqrt{2}$ span the two-body problem in the Chern band, and $A_{\textrm{BZ}}$ is the area of the Brillouin zone. This deviation is due to the breaking of magnetic translation symmetry by the FCI basis.

In analogy to the conventional FQHE, two-body interactions (together with appropriate multi-body generalizations) are therefore well-captured by a Haldane pseudopotential expansion \cite{claassen2015}, which in turn directly entails stability of various FQH many-body trial states at appropriate filling fractions. For the remainder of this paper, the presentation will thus be confined to discussing single-body properties of various ideal isotropic tight-binding host models. Note that given (1) satisfaction of isotropy $\eta^{\mu\nu} = \delta^{\mu\nu}$, (2) the ideal isotropic FCI condition (Eq.\ref{eq:idealIsotropicDropletCondition}), and (3) sufficiently benign Berry curvature nonuniformity over the Brillouin zone, the pseudopotential expansion applies and simple local interactions in the many-body problem suffice to stabilize FQH states at fractional filling.

\section{Ideal FCI criteria}
\subsection*{Bloch eigenfunctions be holomorphic} 

Having discussed the background and motivation for our approach to ideal FCIs via the ideal isotropic FCI condition, we now state specific criteria and constraints for their realization. 


The simplest way to arrive at a lattice model that obeys the ideal isotropic FCI condition (Eq.\ref{eq:idealIsotropicDropletCondition}) employs  {holomorphic coordinates} $z,\bar z = k_x\pm i k_y$. In these coordinates, the metric tensor $g$ takes the form
\begin{equation}
\left(\begin{matrix}
 g_{zz} & g_{z\bar z} \\
 g_{\bar z z} & g_{\bar z \bar z}\\
\end{matrix}\right)=\frac1{2}\left(\begin{matrix}
 & g_{xx}+g_{yy}+2g_{xy} & i(g_{xx}-g_{yy}) \\
 & i(g_{yy}-g_{xx}) & g_{xx}+g_{yy}-2g_{xy}\\
\end{matrix}\right), \\  \label{gij}
\end{equation}
which can be directly derived by comparing the coefficients in $g_{xx}dx^2+ g_{yy}dy^2+2g_{xy}dxdy=g_{zz}dz^2+g_{\bar z\bar z}d\bar z^2 + 2 g_{z\bar z}dzd\bar z$. The ideal isotropic FCI condition is satisfied when $g_{xx}=g_{yy}$ or $g_{z\bar z}=g_{z \bar z}=0$ and $g_{zz}=g_{\bar z \bar z}$, which holds when the normalized Bloch eigenfunction take the form $\varphi=\phi(z)/|\phi(z)|$, where $\phi(z)$ is an unnormalized Bloch eigenfunction which depends only holomorphically\footnote{Equivalently, we can assume purely anti-holomorphic dependence $\phi(\bar z)$, which will result in a very similar conclusion. } on $z=k_x+ik_y$, i.e. is independent of $\bar z = k_x-ik_y$.

In an FCI with $N$ bands, the eigenfunction $\varphi$ has $N$ complex components while satisfying $|\varphi|^2=1$, and thus lives in the complex projective space $\mathbb{CP}^{N-1}\sim \frac{U(N)}{U(N-1)\times U(1)}$. This can also be understood by noticing that it is obtained from the space of nonzero complex $N$-tuples $\mathbb{C}^{N}/\{0\}$ by quotienting out the overall normalization and the $U(1)\sim S^1$ gauge degree of freedom, i.e.
\begin{equation}\mathbb{C}^{N}/\{0\}\sim S^{2N-1}\rightarrow S^{2N-1}/S^1\sim \mathbb{CP}^{N-1} \label{1band}.\end{equation}
For generic, not necessarily normalized nor holomorphic $\phi_n (z,\bar z), $ $n \in 1,2,\dots, N$, $\phi \equiv (\phi_1,\phi_2,\dots,\phi_N)$, the substitution of $\varphi=\phi/|\phi|$ into Eq. \ref{Q} gives expressions for Berry curvature and Fubini-Study metric:
\begin{widetext}
\begin{align}
		F_{xy} &= 2\sum_{n>m}^N \left[ \frac{\left|\phi_m \partial_z \phi_n - \phi_n \partial_z \phi_m \right|^2}{|\phi|^4} - (z\leftrightarrow \bar{z}) \right], \\
	{\rm Tr~}g &= 2\sum^N_{n>m} \left[ \frac{\left|\phi_m \partial_z \phi_n - \phi_n \partial_z \phi_m \right|^2}{|\phi|^4} + (z\leftrightarrow \bar{z}) \right], \\
	g_{xx} - g_{yy} &=  2\sum^N_{n>m} \frac{\left( \phi_m \partial_z \phi_n - \phi_n \partial_z \phi_m \right) \overline{\left( \phi_m \partial_{\bar{z}} \phi_n - \phi_n \partial_{\bar{z}} \phi_m \right)} + {\rm c.c.} }{|\phi|^4}, \\
	g_{xy} &= i\sum^N_{n>m} \frac{\left( \phi_m \partial_z \phi_n - \phi_n \partial_z \phi_m \right) \overline{\left( \phi_m \partial_{\bar{z}} \phi_n - \phi_n \partial_{\bar{z}} \phi_m \right)} - {\rm c.c.} }{|\phi|^4}.
	\label{Fg}
\end{align}
\end{widetext}
Evidently, the ideal \textit{isotropic} FCI condition $\textrm{Tr}  g = F_{xy}$ is satisfied if the $(z\leftrightarrow \bar z)$ term of $F_{xy}$ and $Tr\,g$ vanish. This can be always realized with \textit{holomorphic}  eigenstates $\phi=\phi(z)$, which obey $\partial_{\bar z}\phi=0$. 
Furthermore, the above equations also imply that $\textrm{Tr}  g(k) > |F_{xy}(k)|$ if holomorphicity is violated. The fact that holomorphicity automatically fixes $\textrm{Tr} ~g(k)$ to its lower bound leads to another advantage, since it has been shown that\cite{lee2014lattice} the locality of the parent pseudopotential Hamiltonian of the $1/3$-Laughlin state on the lattice is bounded below by $\int \textrm{Tr} ~g(k)\, d^2k$, at least with the Wannier state mapping\cite{qi2011,lee2013,lee2014lattice}. Hence PP operators on our ideal FCIs have the tendency to be more local, and hence simpler to study theoretically and experimentally.

The parent lattice Hamiltonian that hosts a holomorphic unnormalized Bloch state $\phi(z)$ is not unique. The simplest way to construct such a Hamiltonian such that it also has perfectly flat bands is to start from the complement of the band projector $Q$ (see Section II):
\begin{align}
	H=Q=\mathbb{I}- |\varphi\rangle\langle\varphi|. 
	\label{eq:H0}
\end{align}
Its eigenspectrum hosts a single flat band at energy $\epsilon=0$, and $N-1$ degenerate flat bands at energy $\epsilon=1$. While functions $\phi_m(z)$ are necessarily singular at $C$ points in the BZ, as discussed below, the normalization $\varphi=\phi/|\phi|$ guarantees smoothness of the matrix elements of the Hamiltonian.

Having strongly constrained the ansatz to a class of flat-band models that satisfy the ideal isotropic FCI condition $\textrm{Tr}  g(k) = |F_{xy}(k)|$, we turn, as a next step, to fluctuations of the Berry curvature over the BZ. While pseudopotential decompositions of local interactions for this class of flat-band models have been shown to be robust against weak Berry curvature fluctuations \cite{claassen2015,behrmann2016model}, it is highly desirable to optimally suppress the latter, and bring the FCI model closer to the limit of the isotropic conventional FQH system \cite{PhysRevB.90.165139}. The analogy between FCI and FQH systems is further improved by the ideal isotropic condition $F_{xy}=\textrm{Tr}\,g$ in our construction, since vanishing fluctuations in both of these quantities result in the density algebras for the FCI and the FQH systems agreeing to third order in the long wavelength limit\cite{Roy2012}. This brings us to:

\subsection*{Constraint 1: maximize Berry curvature uniformity}

As previously mentioned, the mean square deviation from uniform Berry flux $\langle(\Delta F)^2\rangle$ has a finitely large lower bound for $N=2$ bands, because it is impossible to map $T^2\rightarrow S^2$ with a constant Jacobian, as can be easily seen by matching grids on these manifolds. However, $\langle(\Delta F)^2\rangle\rightarrow 0$ is theoretically achievable for $N\geq 3$, and in Section \ref{sec:examples} we shall explicitly construct a 3-band model with a $\langle(\Delta F)^2\rangle$ much smaller than all known lattice Hamiltonians with small number of bands and at most next-nearest-neighbor (NNN) hoppings\footnote{Hofstadter-type models can possess very uniform Berry curvature\cite{aoki1996hofstadter,aidelsburger2015measuring}, but they possess a large number of bands due to their large magnetic unit cell.}.

Finally, the Hamiltonian of Eq. (\ref{eq:H0}), while exactly satisfying $\textrm{Tr}~ \mathbf{g} = F_{xy}$ and hosting a perfectly flat band, necessarily introduces long-ranged hoppings. These have been shown to decay exponentially with distance\cite{chen2014impossibility,lee2016band}, but it is nevertheless desirable to deal with simpler local lattice models instead. A key advantage of our proposed choice of Hamiltonian in Eq. \ref{eq:H0} is that it minimizes the mean-square hopping distance of the hopping elements\cite{jian2013momentum}. The proof mirrors that of the $\mathbb{CP}^1$ solution for magnetic Skyrmions\cite{han2010skyrmion}, and relies crucially on the holomorphicity of $\phi(z)$. 
The strategy put forward in this manuscript will be to take models of the form of Eq.(\ref{eq:H0}), which possess hoppings with minimal mean-square distance, and truncate longer-ranged hoppings. In other words, the gist of our approach is to start from an perfectly-isotropic FCI model with known optimal conditions for the pseudopotential decomposition, and deform it to find a compromise between locality and weak violation of the ideal droplet condition $\textrm{Tr}~ g = F_{xy}$, as well as flatness of the band dispersion. We hence arrive at:

\subsection*{Constraint 2: maximize band flatness}
The final condition for an ``ideal'' FCI is that the fractionally filled band has very little dispersion\footnote{Note that we can have lower-lying filled valence bands with significantly nonuniform dispersion, since they do not contribute to the dynamics.} in its energy $\epsilon_1(k)$ relative to its gap to the lowest unoccupied band $\epsilon_2(k)$, so that the interaction term dominates just like in the FQH effect. We can quantify the uniformity of its dispersion via the  {flatness ratio}
\begin{eqnarray}
f=\frac{\text{bandgap}}{\text{bandwidth}}=\frac{\text{min}(\epsilon_2(k))-\text{max}(\epsilon_1(k))}{\text{max}(\epsilon_1(k))-\text{min}(\epsilon_1(k))}
\end{eqnarray}  
where $\epsilon_{1,2}(k)$ are the dispersions of the (partially filled) valence band and lowest conduction band, respectively. Henceforth, the filled eigenstate $\varphi_1$ will be simply denoted as $\varphi$. As discussed in Ref. \onlinecite{lee2016band} and proven in Ref. \onlinecite{chen2014impossibility}, $f$ cannot be arbitrarily large for a topologically nontrivial system with finite-ranged real-space hoppings, at least not for static systems\cite{grushin2014floquet,zhou2014aspects,poudel2015dynamical}. Hence our final goal will be to find a candidate FCI model whose band is as flat as possible - given constraints on the Chern number, lattice geometry, and Berry curvature uniformity.

\section{Recipe for Ideal FCIs}
\label{recipe}
Here we detail the construction of our ideal FCIs, namely, FCIs defined by holomorphic Hamiltonians (Eq. \ref{eq:H0}) such that both Berry curvature and band dispersion are as uniform as possible.

\subsection{Outline of recipe}
We first provide an overview of the main steps of our proposed construction procedure, and then elaborate on them in the following subsections.

\begin{enumerate}
\item Decide on the Chern number $C\geq 2$, and write down an ansatz vector of trial doubly periodic meromorphic functions (elliptic functions) $\phi(z)=(\phi_1(z),...,\phi_{N-1}(z),1)^T$ containing a total of $C$ poles, as described in Sections \ref{sect:instanton} and \ref{sect:berry}. 
The Hamiltonian, in its diagonal basis, has matrix elements (Eq. \ref{eq:H0}), 
\begin{equation}
H_{mn}=\delta_{mn}- \frac{\phi_m(z) \phi^*_n(\bar z)}{|\phi(z)||\phi(z)|}.
\label{Hmn}
\end{equation}
with $z=k_x+ik_y$ and perfectly flat lowest band with Chern number $C$ and eigenfunction $\varphi=\phi/|\phi|$. Note that this procedure only specifies $H$ up to its eigenbasis; one can freely rotate it to any other basis as desired. 

\item Minimize the Berry curvature nonuniformity $\langle (\Delta F)^2\rangle$ (Eqs. \ref{dFxy} and \ref{Fg}) by varying the ansatz $\phi$.  Key to the elegance of this minimization is the small number of parameters, since elliptic functions are uniquely determined by their $C$ poles, each of which is characterized only by position, order and value of residue\footnote{The difference of two of two such functions have no poles and must thus be constant by Liouville's theorem.}.

\item Truncate longer-ranged hoppings of the resultant Hamiltonian (Eq. \ref{Hmn}), which serves to balance the need for a simple effective local tight-binding model with corresponding deviations from the ideal quantum geometry of the Chern band.

\item  Since a real-space truncation reintroduces a parasitic dispersion to the Chern band (Sect.\ref{sec:complex}), the resultant band may be optimally flattened by adding a suitable counteracting ``identity'' dispersion term that does not affect the quantum geometry. This last step of flattening is most efficient when the model parameters are tuned to maximize the imaginary gap, which controls the decay rate of the dispersive terms produced by the truncation\cite{lee2016band}. In any case, the truncation is unlikely to compromise the band flatness significantly, because the holomorphicity of $\varphi(k_x+ik_y)$ in our Hamiltonian already serves to minimize the mean-square hopping distance even before the truncation\cite{jian2013momentum}. 




\end{enumerate}

Note that the optimizations in steps 2 and 4 may not necessarily lead to the same optimal final solution, and an appropriate weight should be decided based on the relative importance of Berry curvature and band dispersion uniformity.
All steps will be explicitly elaborated on in the following subsections.

\begin{figure}[h]
 \includegraphics[width=0.49\linewidth]{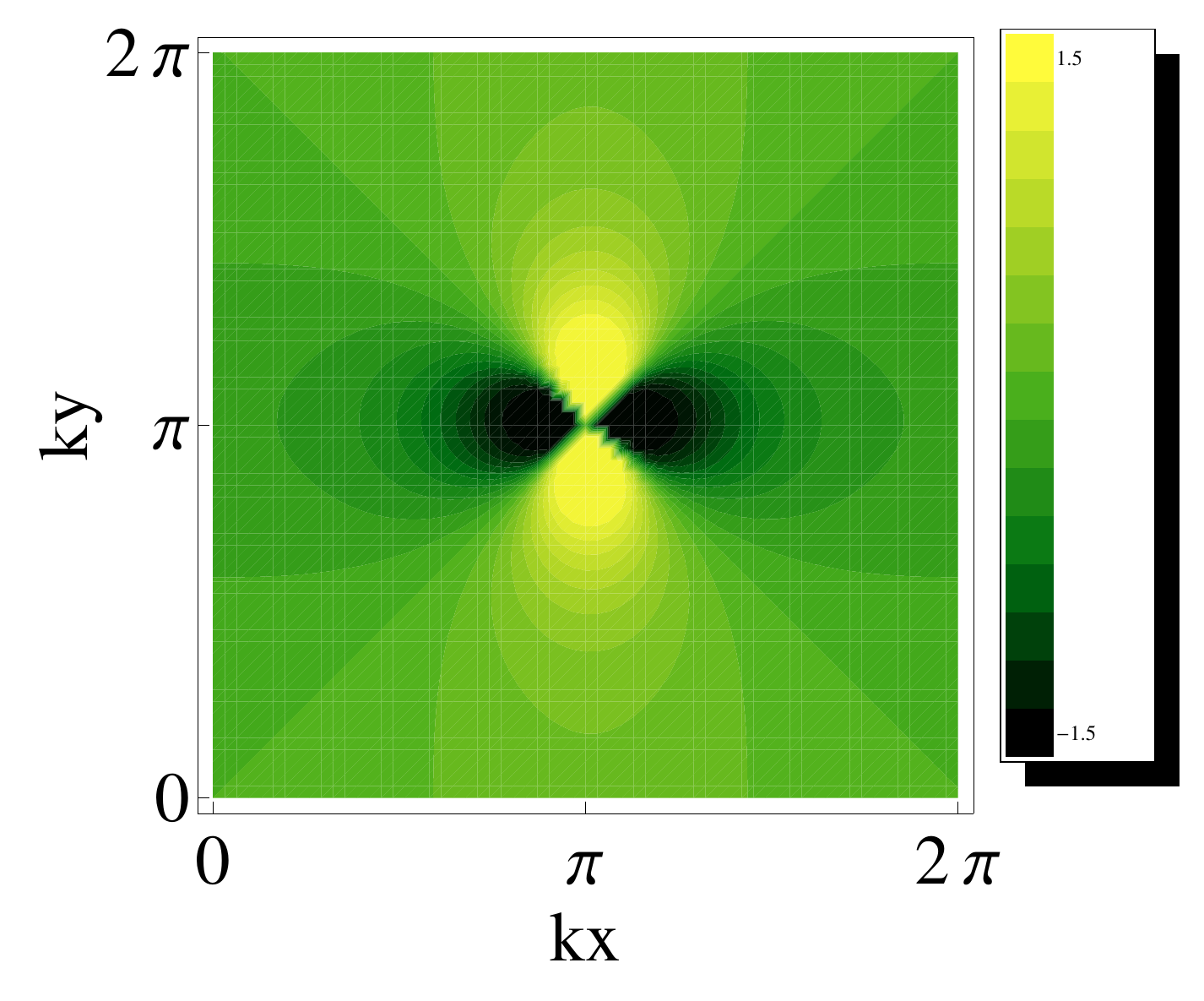}
 \includegraphics[width=0.49\linewidth]{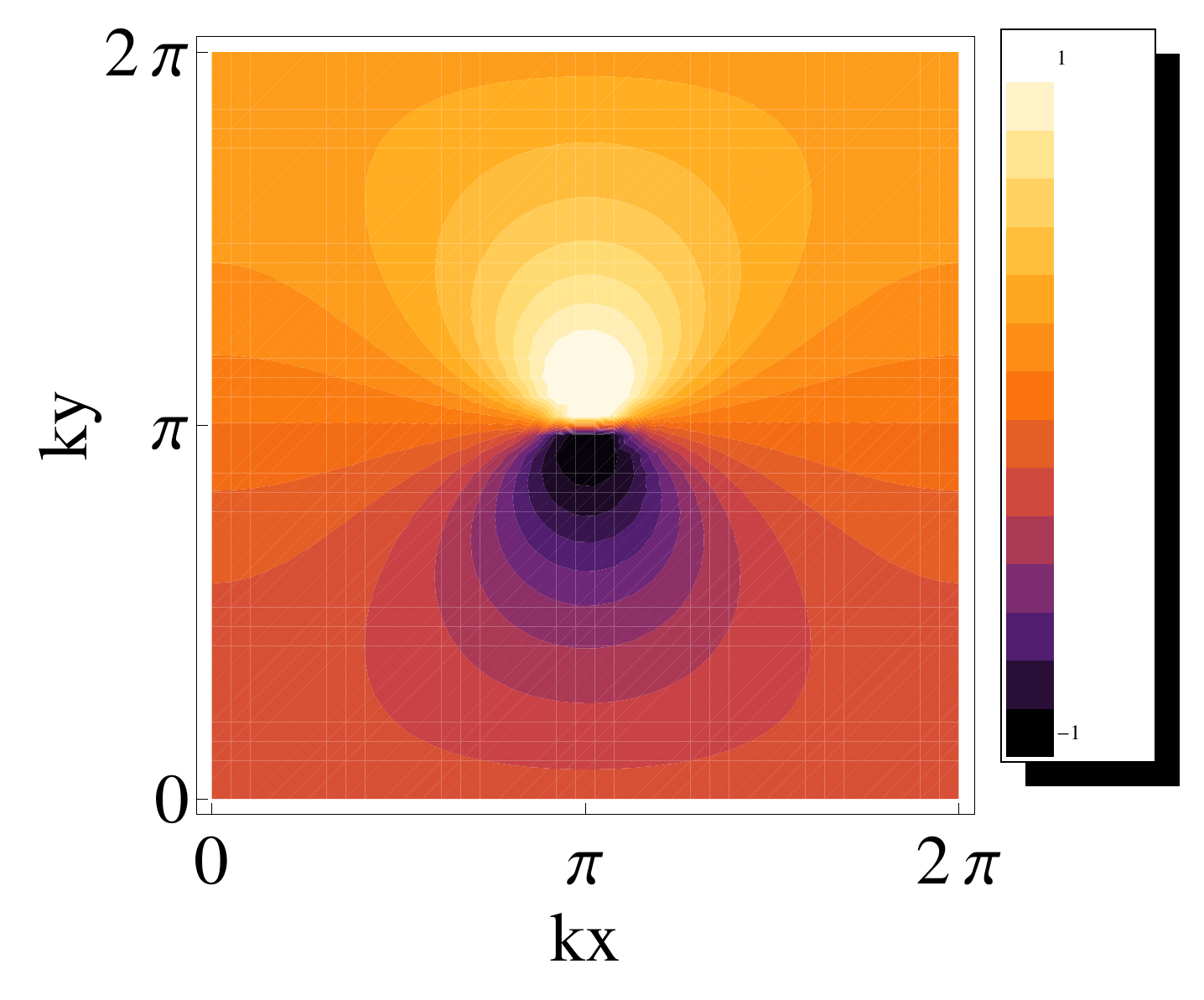}
 \includegraphics[width=0.49\linewidth]{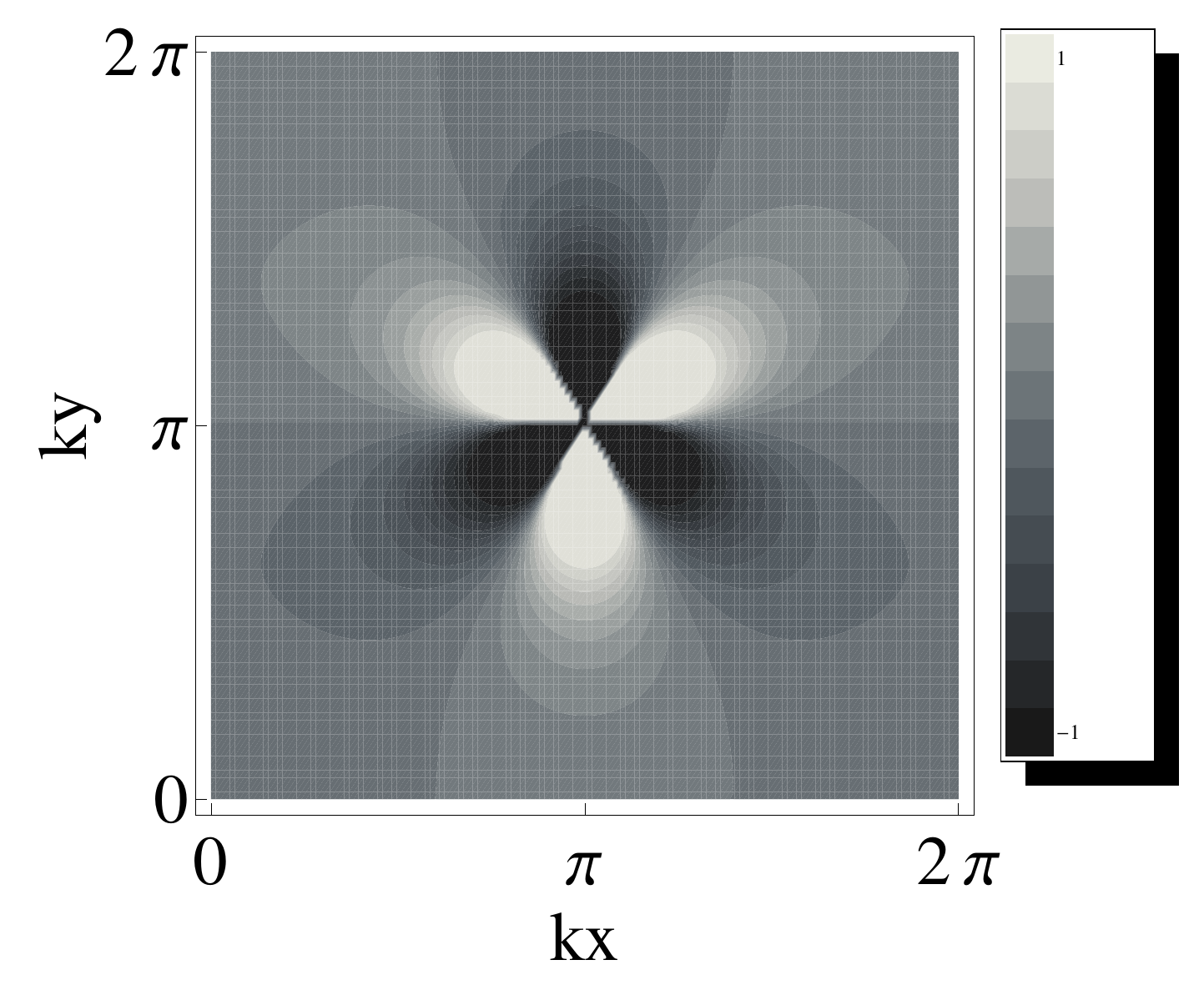}
 \includegraphics[width=0.49\linewidth]{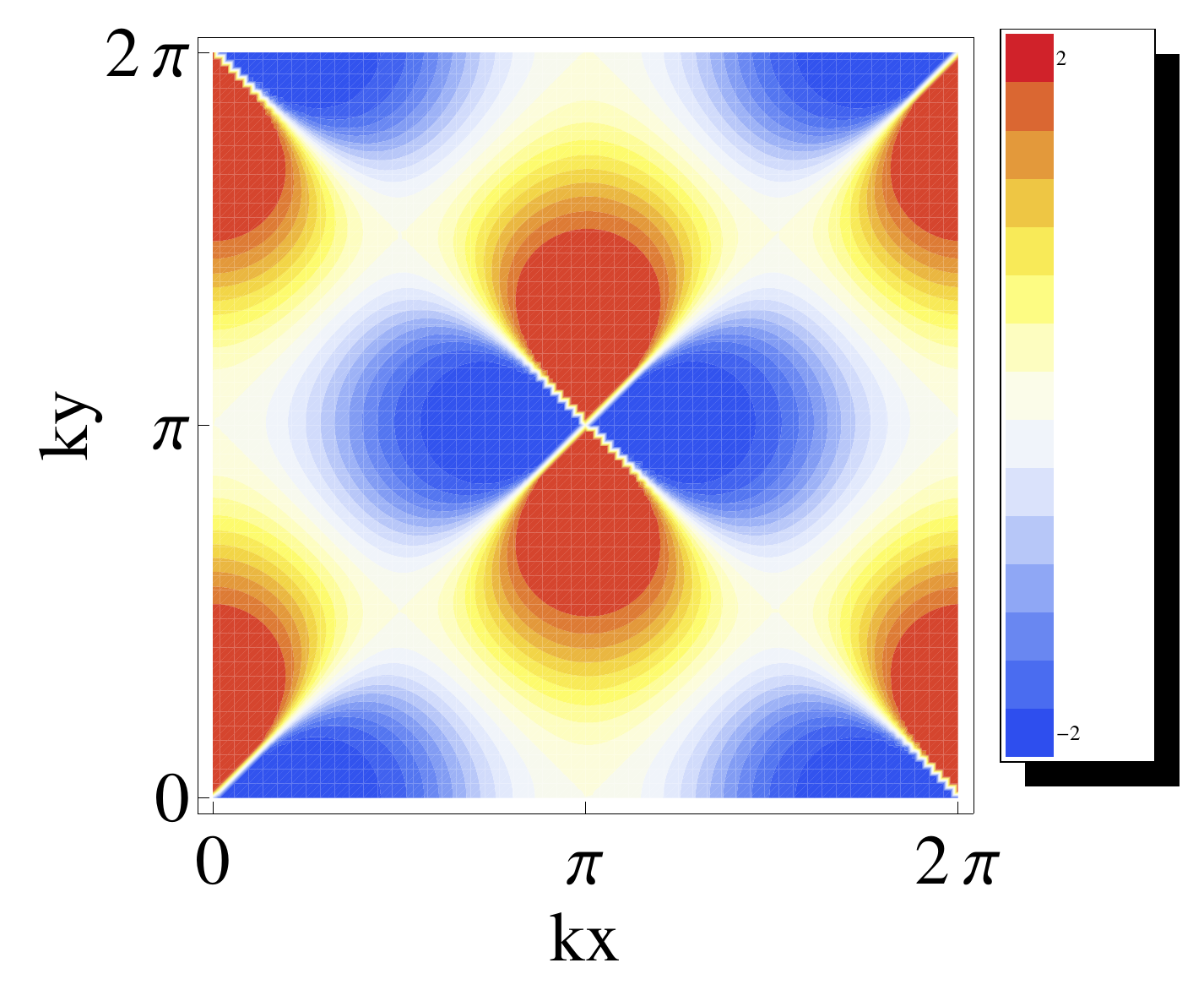}
\caption{(Color Online) The real parts of illustrative elliptic functions used in our construction of ideal FCI Bloch bands. Top Left) The Weierstrass elliptic function $W(z+\pi(1+i))$ (Eq. \ref{phizconstruction}) with a double pole at the origin. Top Right) Its antiderivative, the Weierstrass Zeta function $\zeta(z+\pi(1+i))$ with a single pole at the origin. Note that it is not doubly periodic, as forbidden for meromorphic functions with only one pole; however, linear superpositions of pairs of such Weierstrass Zeta functions can be doubly periodic. Bottom left) $W'(z+\pi (1+i))$ with a third order pole at the origin. Bottom Right) The function $(W'(z)/W(z))^2$ used in the Chern number $4$ model in Eq. \ref{model3}; it inherits its set of two double poles from the poles of $W'(z)$ and the zeroes of $W(z)$.}
\end{figure}

\subsection{Implementation of the recipe for ideal FCI construction}
\label{sect:instanton}

The ansatz Hamiltonian is given by $H_{mn}=\delta_{mn}- \varphi_m \varphi^*_n$ (Eq.~\ref{Hmn}), which hosts a lowest band of eigenenergy $0$ spanned by the Bloch eigenstate $\varphi=\phi(z)/|\phi(z)|$, $z=k_x+ik_y$. Although this is a perfectly flat band by definition, some dispersion will be introduced upon real-space truncation to a more realistic Hamiltonian  (Sect.\ref{sec:complex}). We can fix the overall phase and normalization of $\phi(z)$ by the parametrization
\begin{equation}
\phi(z)=(\phi_1(z),\phi_2(z),...,\phi_{N-1}(z),\,1)^T,
\label{phiz}
\end{equation}
i.e. $\phi_N=1$. Since $k$ lives in the periodic BZ, each $\phi_i$ must be a doubly periodic (elliptic) meromorphic function, i.e. a function in $z=k_x+ik_y$ that are holomorphic except at isolated poles. In general, the $\phi_i$'s should all have the same periodic unit cell, which is determined by symmetry considerations of the desired lattice symmetry of the Hamiltonian $H$ in real-space.

For now, we shall assume the simplest possible symmetry where the unit cell is a square, so that each $\phi_j$ (denoted as $\phi$ for simplicity from now on) satisfy $\phi(z)=\phi(z+2\pi)$ and $\phi(z)=\phi(z+2\pi i)$. Functions with these properties are most easily expressed\footnote{Equivalently, such doubly periodic functions are expressible in terms of rational combinations of $W(z)$ and $W'(z)$.} as linear combinations of 
Weierstrass Zeta functions $\zeta(z)$ and their derivatives, the Weierstrass elliptic functions $\zeta'(z)=-W(z)=-\sum_{p,q\in \mathbb{Z}/\{0\}}\left[(z+2\pi(p+qi))^{-2}-(2\pi(p+qi))^{-2}\right]$:
\begin{equation}
	\phi(z)=\sum_j a_j \zeta(z-b_j) + f(\{W(z-b'_j)\}) + c,
	\label{phizconstruction}
\end{equation}
where $a_j,b_j$, and $c$ are tunable parameters, and $f(\cdot)$ is a holomorphic function. Here, the role of the Zeta functions is to place single poles at positions $b_j$, with constraint $\sum_ja_j=0$ ensuring that the resulting function is elliptic. The fact that $\zeta(z)$ has a single pole at $z=0$ can be deduced from the pole structure of $W(z)$ at $z=0$, where\footnote{The difference in the parentheses ensures that the sum does not blow up at all $z$ except $z=0$, where there is a double pole $\sim \frac1{z^2}$.} $W(z)\sim \frac1{z^2}$. The role of $f(\{W(z-b'_j)\})$ is to place higher-order poles at positions $b'_j$. For instance, $W(z)$ and $W^2(z)$ place a double and quadruple pole at the $\Gamma$ point, respectively.
As a quintessential elliptic function, $W(z)$ has also appeared in innumerable other theoretical contexts such as combinations, number theory, conformal field theory, and topology\cite{conway1979monstrous,gannon2006moonshine,cheng2011rademacher}. 

To relate the pole structure of the $\phi(z)$'s with its Chern number, note that 
\begin{eqnarray}
C&=&\frac1{2\pi}\int_{BZ} \left(\partial_x A_y-\partial_y A_x\right)\notag\\
&=& \frac1{4\pi}\left(\oint_{\partial BZ}+\sum_j \oint_{\Gamma_j}\right)\left(A_{\bar z}dz + A_z d \bar z \right)\notag\\
&=& -\frac1{2\pi i}\sum_j \oint_{\Gamma_j}\frac{\sum_l^{N-1}|\phi_l(z)|^2\partial_z \log \phi_l(z)}{1+\sum_l^{N-1}|\phi_l(z)|^2}dz\notag\\
&=&  -\frac1{2\pi i}\sum_j\sum_l^{N-1} \oint_{\Gamma_j}\partial_z \log\phi_l(z)\notag\\
&=& \sum_j R_{j},
\label{chern}
\end{eqnarray}
with $A_j=-i\langle \varphi|\partial_j\varphi\rangle$ the Berry connection, and $R_{j}$ the order of the $j^{th}$ pole among functions $\phi_1(z),...,\phi_{N-1}(z)$. The complex Stokes' theorem was used from the first to the second line, where the $\Gamma_j$'s are infinitesimal contours around all the poles and zeros of $A_z$ and $A_{\bar z}$. The $\oint_{\partial BZ}$ term disappears due to equal but opposite contributions from the opposite sides of the BZ. 
As the $\Gamma_j$ loops are infinitesimally small, contributions due to the zeroes and nondivergent points of $\phi_l(z)$ are negligible. As such we are left with the windings of $\partial_z \log\phi_l(z)$ in the fourth line, which are just the $R_j$'s. 

Hence the task of designing a system with Chern number $C$ reduces to that of deciding on $N \leq C$ poles with positions $b_1,...,b_j$ and residues $R_j$ such that $\sum R_j = C$. The simplest route involves using only  $C$ poles of order one, which merely involves choosing the parameters $a_j$ and $b_j$ in $\phi(z)=\sum_ja_j\zeta(z-b_j)$, such that $\sum_j a_j=0$. More generally, one can utilize $f(\cdot)$ in Eq. \ref{phizconstruction} to conformally map $W(z)$ to one of a desired pole structure. 
Note that $\zeta(z)$ has only one single pole, and as proven in Appendix \ref{sect:elliptic} cannot be doubly periodic. However, the combination $\phi(z)$ is \textit{indeed} double periodic as long as $\sum_j a_j=0$. This implies that we need to use at least two Weierstrass Zeta functions, or resort to Weiertrass elliptic functions, constraining the Chern number to $C \geq 2$. There are no constraints in the maximum number of orbitals in the resulting tight-binding model, however, even if the multiple components of $\phi(z)$ have identical poles. Our approach is closer to the currently used FCI models than it might appear at this stage. As we shall see in the following, various simple optimal configurations of $\phi(z)$ reproduce, after real-space truncation, flatband models that have already surfaced in the existing literature. In our approach, however, the search for much more complicated ideal FCI models with higher Chern number and number of bands does not involve much bigger effort.


\subsection{Models for a specific number of bands $N$}
\subsubsection{$N=2$ bands}
\label{N2bands}
The simplest $2$-band case (with $1$ occupied band) deserves special mentioning for pedagogical and aesthetic reasons.  
A $2 \times 2$ Hamiltonian $H=\sigma\cdot \vec d(k)$ is a map from the torus to the Bloch sphere $T^2 \rightarrow S^2\sim \mathbb{CP}^1/U(1)$. 
This mapping can be broken down into the sequence $T^2\rightarrow \mathbb{CP}^1\cup \infty\rightarrow S^2$, where the first map is simply given by the elliptic function $\phi_1(z)$, $z\in T^2$, and the second map the stereographic projection onto the Bloch sphere
\begin{align}
\hat d_1&= \frac{2Re(\phi_1)}{1+|\phi_1|^2},\\
 \hat d_2&= \frac{-2Im(\phi_1)}{1+|\phi_1|^2},\\
 \hat d_3&= \frac{1-|\phi_1|^2}{1+|\phi_1|^2},
\label{d}
\end{align}
i.e., $\phi(z)=(\phi_1(z),1)^T\propto (\hat d_1-i\hat d_2,1+\hat d_3)^T$. It is easy to check that $|\vec d|^2=1$, and that the zeroes/poles of $\phi_1(z)$ correspond to points where $\hat d_3=\pm 1$. Due to Picard's little theorem, $\phi_1$ ranges over all of $\mathbb{C}$ except for at most one point\footnote{Suppose not, that $\phi_1$ omits two points, i.e. $\phi_1:\mathbb{C}\rightarrow M_{0,1}$, the twice punctured sphere. It is known that $\mathbb{H}$, the hyperbolic space, is an universal cover for $M_{0,1}$. Hence there exist a surjective map $f_1:\mathbb{H}\rightarrow M_{0,1}$. We also know that we can find a map $f_2:\mathbb{C}\rightarrow \mathbb{H}$ such that $\phi_1=f_1\circ f_2$. This is because both $\mathbb{C}$ and $\mathbb{H}$ are both simply connected and hence related by homotopy. But that fact that $\mathbb{H}$ is simply connected means that $f_2$ cannot have poles, and must be constant by Lioville's theorem. Hence $\phi_1$ must be constant as well, i.e. a non-constant $\phi_1$ cannot omit more than one point.}, which is of measure zero. Hence the number of poles measures the number of times $f$ wraps around a chosen point, i.e. the degree of the map. This provides a geometric illustration for Eq. \ref{chern}.

The geometric meanings of $\textrm{Tr}  g$ and $F_{xy}$ become even clearer if we express $\vec d$ in terms of the $(\theta,\lambda)$ coordinates on the Bloch sphere: $\vec d=(\sin\theta\cos\lambda,\sin\theta\sin\lambda,\cos\theta)^T$. With this parametrization,  the 2-component eigenstate takes the spinor form $\varphi=\left(\cos\frac{\theta}{2},e^{i\lambda}\sin\frac{\theta}{2}\right)^T$ for which $\theta$ is periodic in $4\pi$, not $2\pi$. 
From Eq. \ref{Q}, we obtain 
\begin{eqnarray}
g_{ij}&=&\frac1{4}\partial_i \hat d \cdot \partial_j \hat d\notag\\
&=& \frac1{4}\left[(\partial_i\theta)(\partial_j\theta) + \sin^2\theta (\partial_i\lambda)(\partial_j\lambda)\right],
\label{gij}
\end{eqnarray} 
which expresses the quantum distance between 2-component spinors as the geometric distance between their images on the Bloch sphere. From Eq. \ref{gij}, one easily obtains
\begin{eqnarray}
\textrm{Tr}\,g &=& \frac{\sum_i  |\partial_i  \vec d|^2}{4|\vec d|^2}-\left(\frac{\sum_i \partial_i |\vec d|^2}{4|\vec d|^2}\right)^2\notag\\
&=& \frac{1}{4}\sum_i |\partial_i  \hat d|^2,
\label{trq2}
\end{eqnarray}
where $i =x,y$. The RHS of the first line disappears since $\vec d$ is assumed to be already normalized. Hence $\textrm{Tr} \, g$ is $1/4$ of the sum of the squared magnitudes of $\partial_i \hat d$ in both $i =k_x,k_y$ directions, which are infinitesimal changes in $\hat d$ due to displacements in the BZ.

The Berry curvature $F_{xy}$ also possesses a familiar geometric expression in terms of the Jacobian of the map $T^2\rightarrow S^2$:
\begin{eqnarray}
F_{xy} &=& \frac1{2}\frac{\vec d\cdot(\partial_x \vec d \times\partial_y \vec d)}{|\vec d|^3}\notag\\
&=& \frac1{2}\hat d\cdot(\partial_x \hat d \times \partial_y\hat d)\notag\\
&=&\frac1{2}\sin\theta(\partial_x\theta\,\partial_y \lambda-\partial_y\theta\,\partial_x \lambda)\notag\\
&=& \frac1{2}\sin\theta d\theta\wedge d\lambda,
\label{Fxy2}
\end{eqnarray}
which is half the area swept out on the Bloch sphere above the infinitesimal area element on the BZ. 

The ideal isotropic FCI condition requires that $g_{xx}=g_{yy}$, and that $g_{xy}=g_{yx}=0$. This simply means that $|\partial_x \hat d|=|\partial_y \hat d|$ and that $\partial_x \hat d \cdot \partial_y \hat d=0$. Combining these constraints with Eqs. \ref{trq2} and \ref{Fxy2}, the ideal isotropic FCI condition $Tr\, g = F_{xy}$ is recast geometrically as
\begin{align}
\frac1{4}\left(|\partial_x \hat d|^2+|\partial_y \hat d|^2\right)&=\frac1{2}|\hat d||\partial_x \hat d||\partial_y \hat d|\sin \theta_d\notag\\
\Rightarrow \theta_d &= \frac{\pi}{2}
\label{dtriad}
\end{align}
where $\theta_d$ is the angle between $\hat d$ and the plane spanned by $\partial_x \hat d$ and $\partial_y \hat d$. In other words, our ideal 2-band FCIs have $\vec d$, $\partial_x \vec d$ and $\partial_y \vec d$ constantly forming a  {mutually orthogonal} triad. 

This orthogonality can also be alternatively derived from the second line of Eq. \ref{gij}. $g_{xy}=0$ gives $\sin^2\lambda=-(\partial_x\theta)(\partial_y\theta)/((\partial_x\lambda)(\partial_y\lambda))$ which, upon substitution into the equality $g_{xx}=g_{yy}$, yields
\begin{equation}
(\partial_x\theta)(\partial_x\lambda)+(\partial_y\theta)(\partial_y\lambda)=\nabla_k \theta\cdot \nabla_k\lambda=0.
\label{dtriad2}
\end{equation}
Together with the normalization constraint $\nabla_k (\hat d\cdot \hat d)=0$, we arrive at the equivalent conclusion that the $\vec d$, $\nabla_k \theta$ and $\nabla_k\lambda$ of our ideal FCIs form a mutually orthogonal triad.





\subsubsection{$N\geq 3$-bands}
From Eqs. \ref{Fg}, we have seen that the ideal isotropic FCI condition $F_{xy} = \textrm{Tr}\, g$ holds if $\phi_j$ are elliptic functions (holomorphic in $z$ and doubly periodic). In this case, the Fubini-Study metric is conformally flat and we have
\begin{align}
	F_{xy} = Tr\, g =2\sum_{n>m}^N \frac{\left|\phi_m \partial_z \phi_n - \phi_n \partial_z \phi_m \right|^2}{|\phi|^4},
\end{align}
which, upon setting $\phi_N=1$, reduces to 
\begin{equation} 
\textrm{Tr} \,g,F_{xy} = \frac{2(|\partial_z \phi_1|^2\pm|\partial_{\bar z} \phi_1|^2)}{(1+|\phi_1|^2)^2}
\label{Trg2}
\end{equation}
for $N=2$ and
\begin{equation} 
F_{xy} = 2\frac{|\partial_z \phi_1|^2+|\partial_z \phi_2|^2+|\phi_1 \partial_z\phi_2-\phi_2\partial_z \phi_1|^2}{(1+|\phi_1|^2+|\phi_2|^2)^2}-(z\leftrightarrow \bar z)
\end{equation}
for $N=3$. Explicit construction schemes for 3-band Chern bands exist\cite{lee2015arbitrary}, but they typically produce very inhomogeneous Berry curvature unlike the examples that we will present in Sect. \ref{sec:examples}. Further details on the geometric parameterization of the $N=3$ case are given in Appendix~\ref{sect:Trg}.

In all of the cases with a holomorphic ansatz, the Berry curvature can also be expressed as the Laplacian of a K\"ahler potential: $F_{xy}=\partial_z\partial_{\bar z} \left[\frac1{2}\log\sum_i|\phi_i(z)|^2\right]$ (Appendix \ref{sect:berry}). The problem of finding a uniform Berry curvature is thus reduced to that of finding a Laplacianless scalar function on a torus, which possesses known solutions that can be used as starting points (Eq. \ref{Laplacian2}).

\subsection{Truncation errors and complex singularities}
\label{sec:complex}
Finally, we mention how the errors from truncation to a local Hamiltonian may be estimated from the complex analytic properties of $H(k_x+ik_y)$. 
Our truncation approach has already been already systematically developed in Ref. \onlinecite{lee2016band}, so here we shall just provide a brief overview. Further details on the complex analytic properties of elliptic functions, in particular the Weierstrass elliptic function $W(z)$, are found in Appendix \ref{sect:holoham}.

In essence, a real-space truncation of a Hamiltonian $H(k_x,k_y)$ is a Fourier series truncation in both the $k_x$ and $k_y$ directions. It was shown, e.g. in Refs.~\onlinecite{lee2016band} and \onlinecite{he2001exponential}, that for each direction $k=k_x$ or $k_y$, the $l^{th}$ Fourier coefficient $a_l\propto \int e^{ilk}H(k)dk$ decays like $|a_l|\sim e^{-gl}$ asymptotically. Here $g=|Im(k_0)|$, where $k_0$ is the closest singularity of the energy band from the real axis. Such singularities can be points where the eigenenergies diverge or, more typically, where energy bands touch. $g$ is also somewhat confusingly called the ``imaginary gap'' in the literature, since it quantifies how far a gapped physical Hamiltonian is from gap closure. 

It follows that the flatness ratio of the ideal FCIs from our approach should scale like\footnote{This is rigorously justified in the appendix of Ref. \onlinecite{lee2016band}, though with possible exceptions.} 
\begin{equation}
f\sim e^{-l_xg_x - l_y g_y},
\end{equation}
 where $l_i$ is the range of hoppings terms in the $i^{th}$-direction  {after} truncation. Consider for instance a truncation that keeps only the nearest-neighbor and next-nearest-neighbor (NN and NNN) hoppings on the square lattice. They correspond to lattice displacements $(\pm 1 ,0),(0,\pm 1)$ and $(\pm 1,\pm 1)$, i.e. $l_x=l_y=1$. We will also have $g_x=g_y$ if the Hamiltonian is symmetric under $(x\leftrightarrow y)$. 

Calculational examples of $g$ for $H(k_x+ik_y)$ will be presented in Appendix \ref{sect:holoham}. Care has to be taken in their analytic continuation, which is distinct from the holomorphic identification $z=k_x+ik_y$. 

\newcolumntype{C}{>{\centering\arraybackslash}m{10em}}
\begin{table*}\sffamily
\begin{tabular}{l*5{C}@{}}
\toprule
MODEL & $C=-2$ (Eq. \ref{model1}) & $C=-2$ (Eq. \ref{model2}) & $C=4$ (Eq. \ref{model3}) & $C=-1$ Checkerboard & $C=-1$ Honeycomb \\ 
\midrule
Berry $F_{xy}$ & \includegraphics[width=10em]{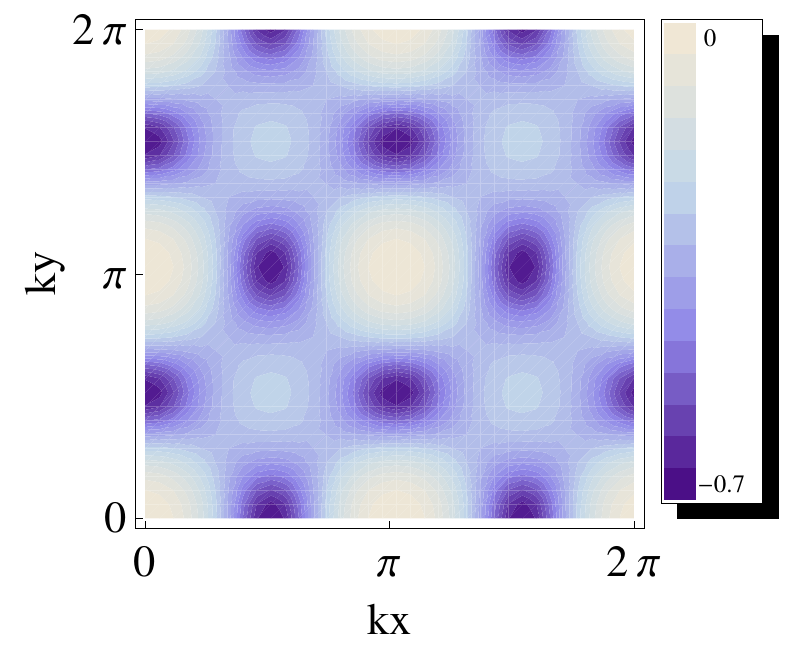} & \includegraphics[width=10em]{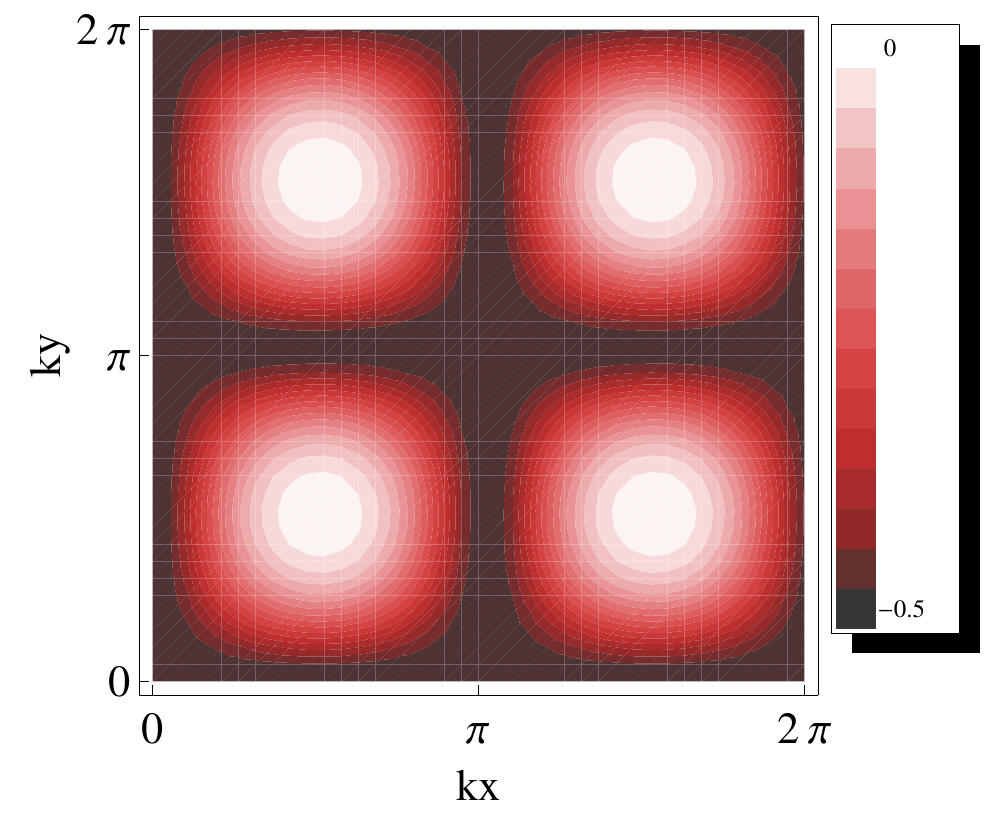} & \includegraphics[width=10em]{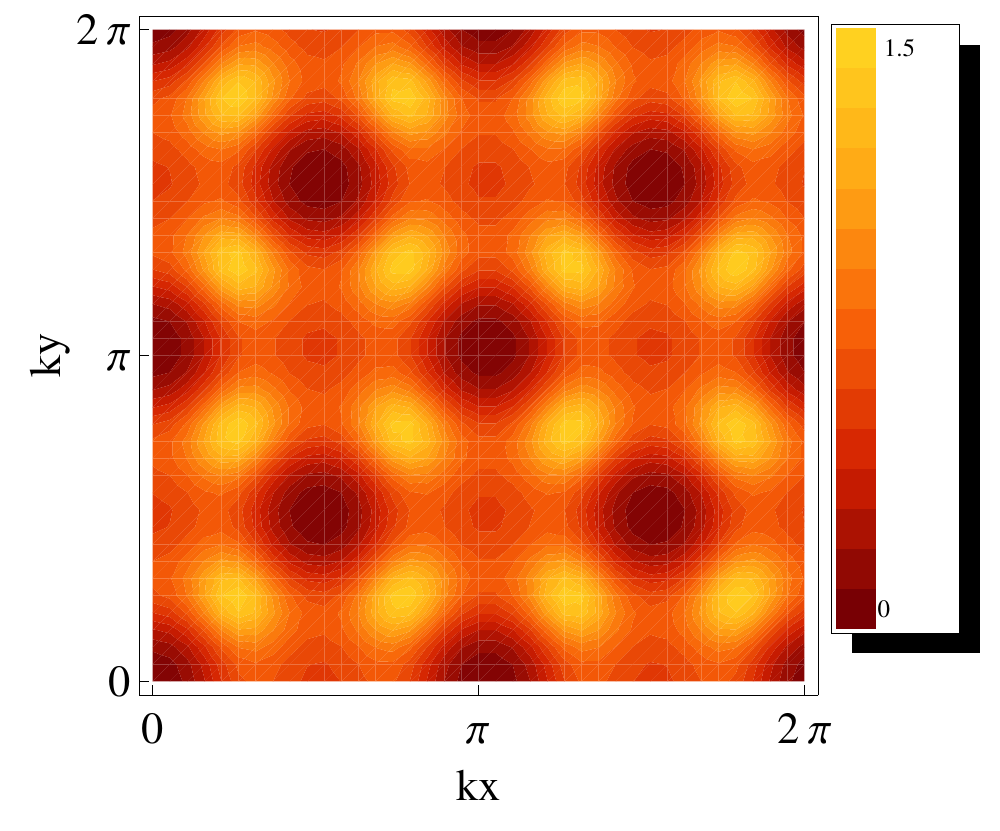} & \includegraphics[width=10em]{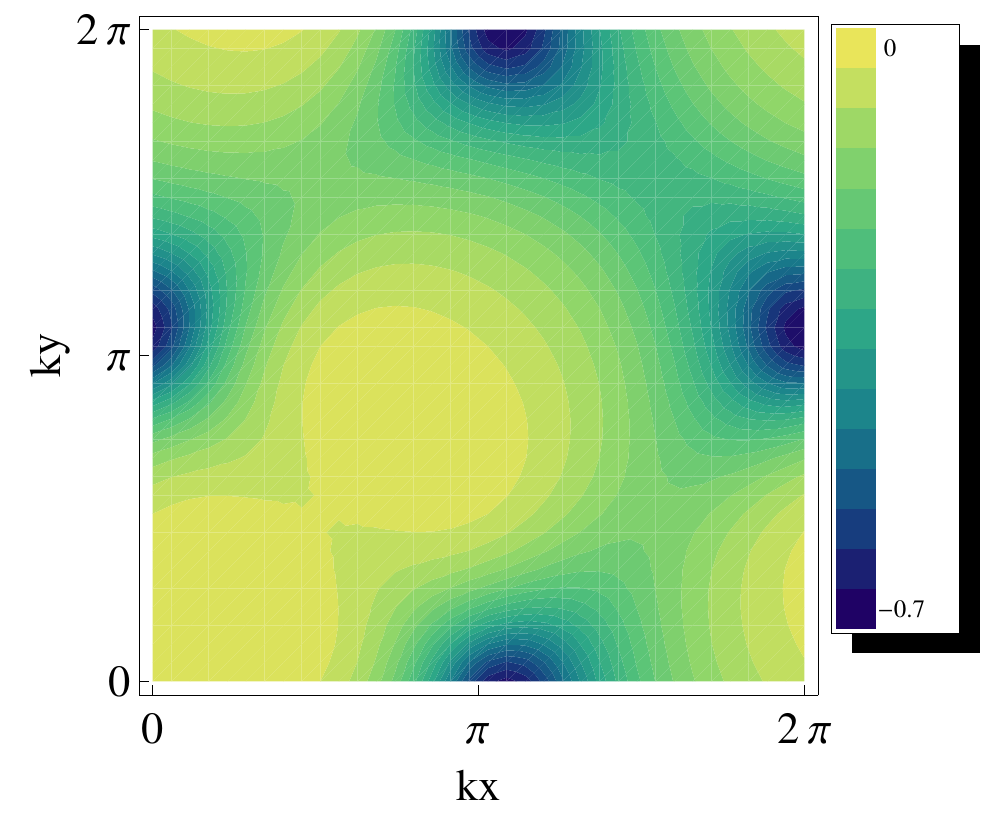} & \includegraphics[width=10em]{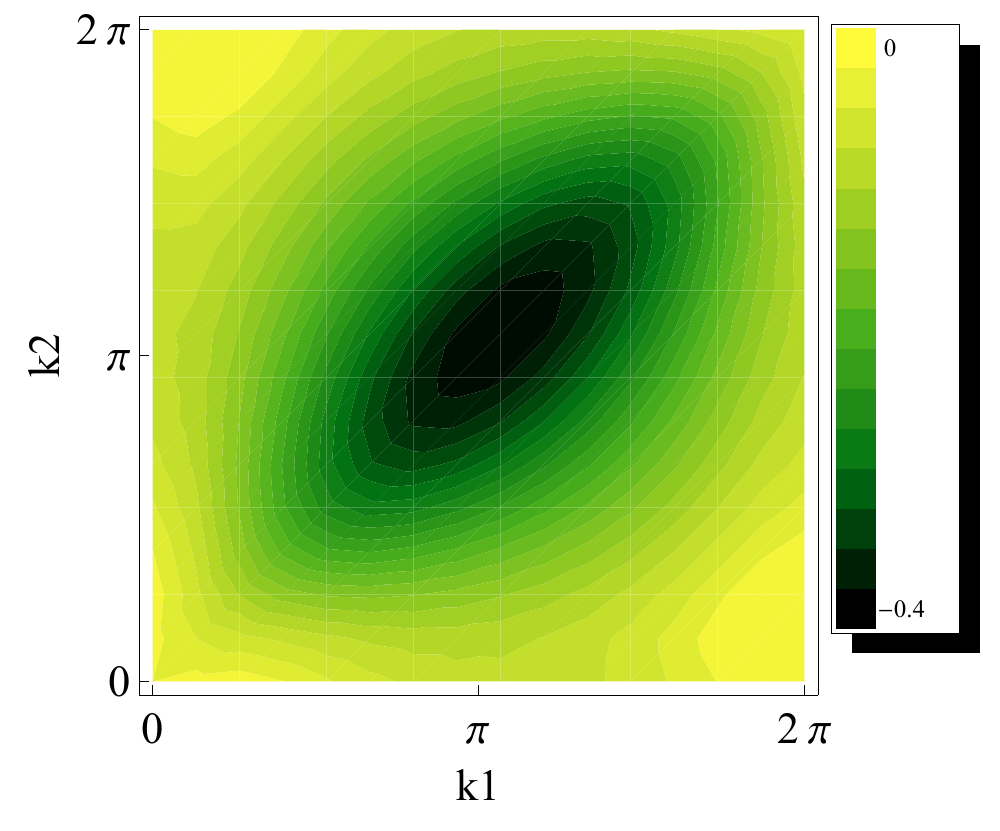} \\ 
d-vector & \includegraphics[width=10em]{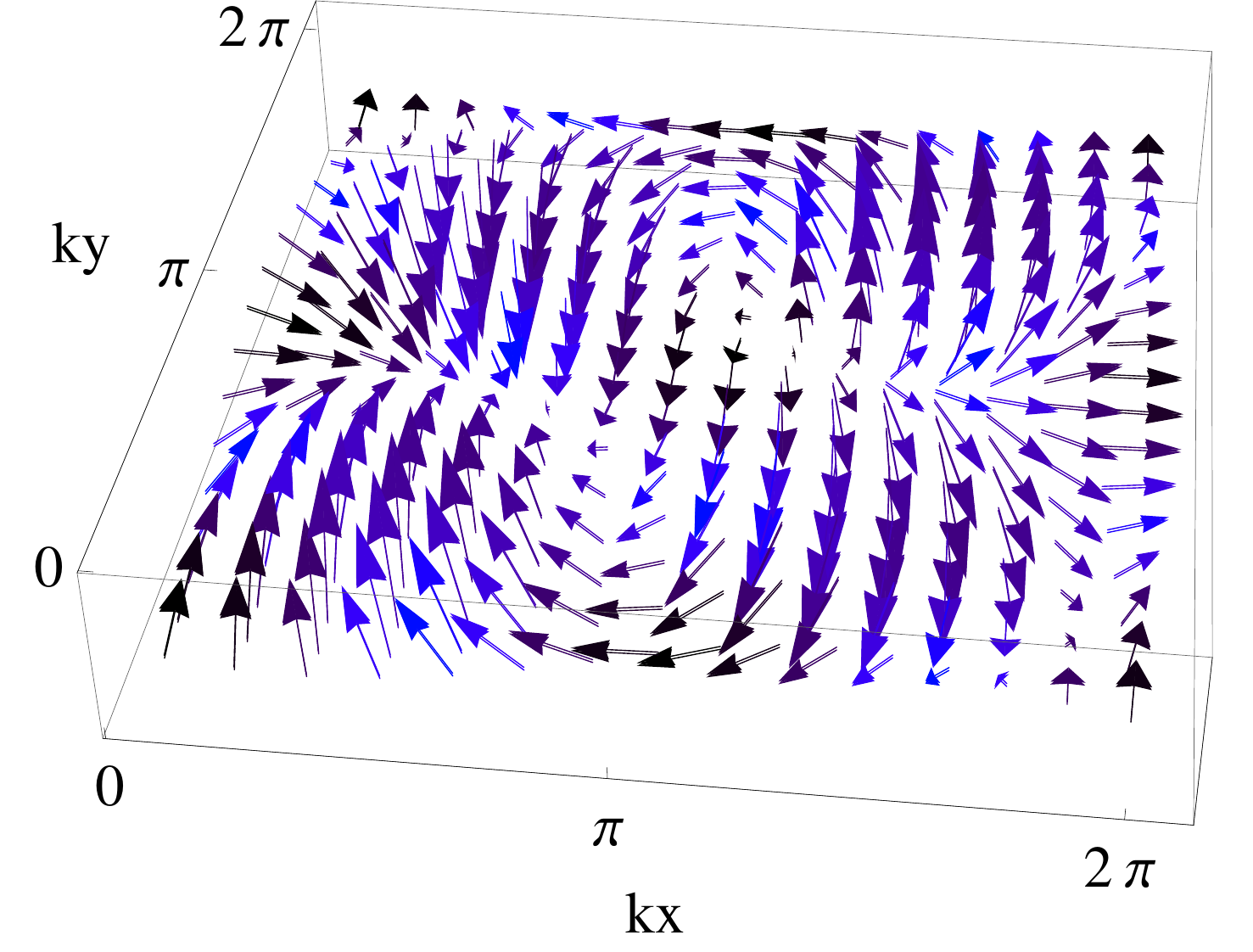} & \includegraphics[width=10em]{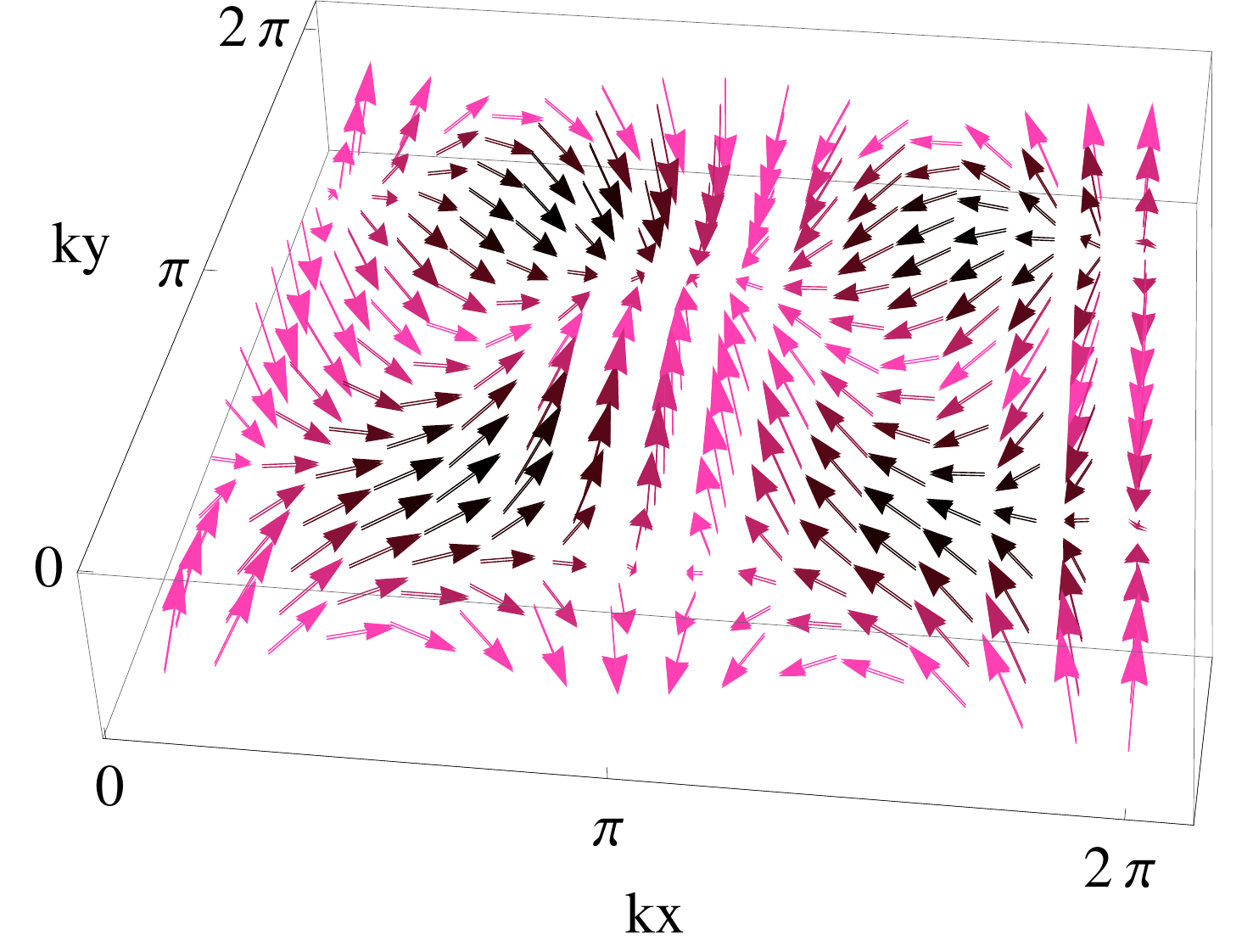} & \includegraphics[width=10em]{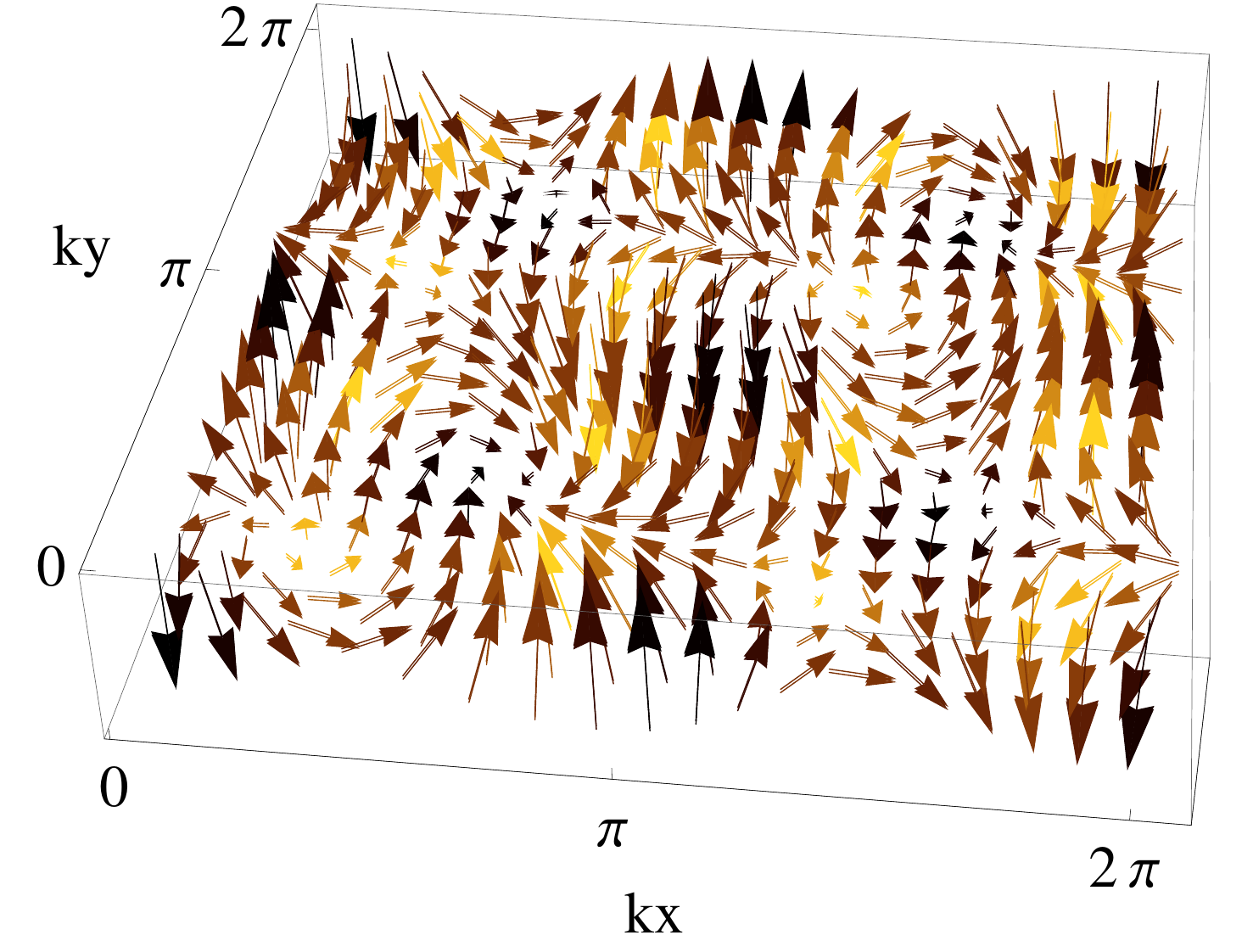} & \includegraphics[width=10em]{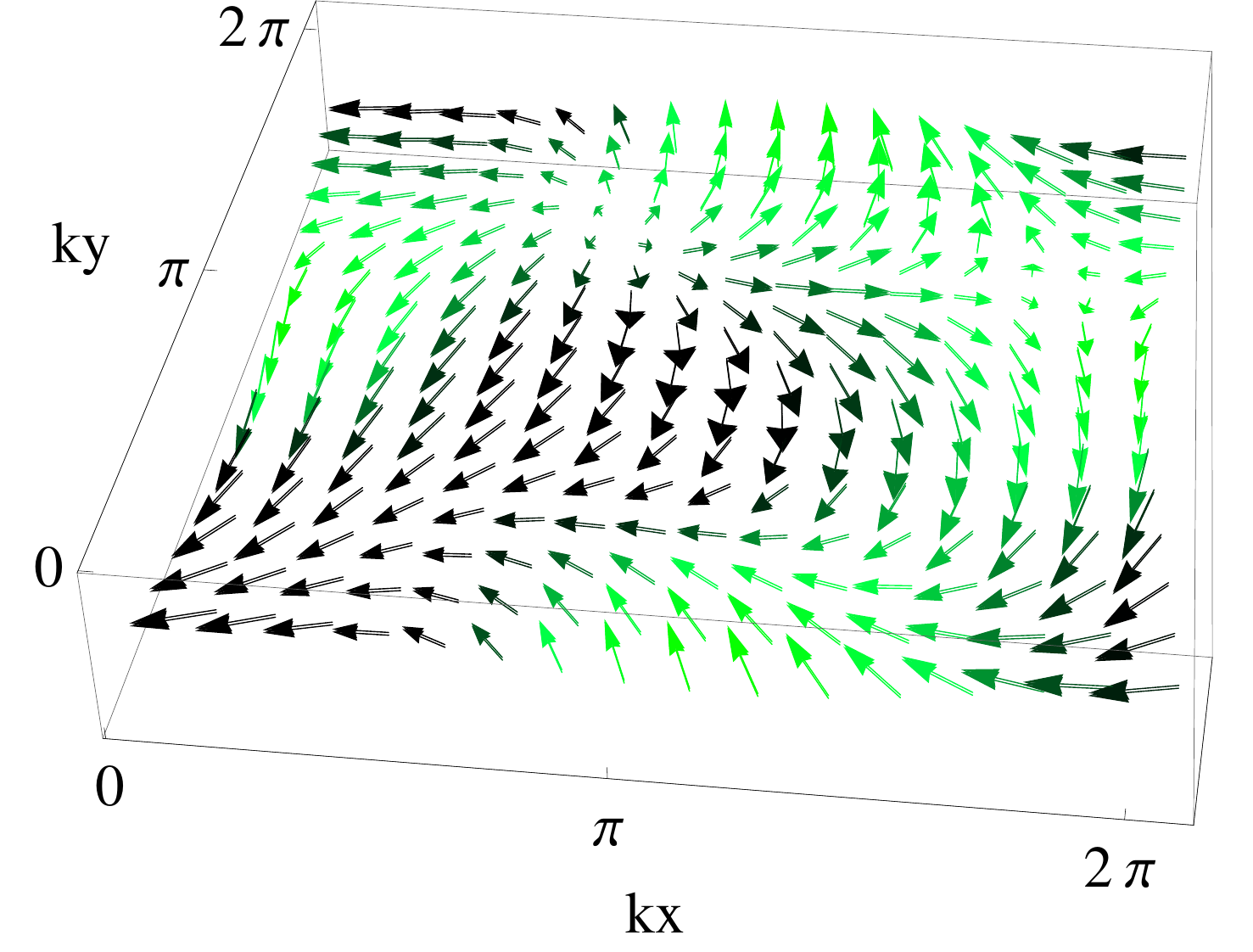} & \includegraphics[width=10em]{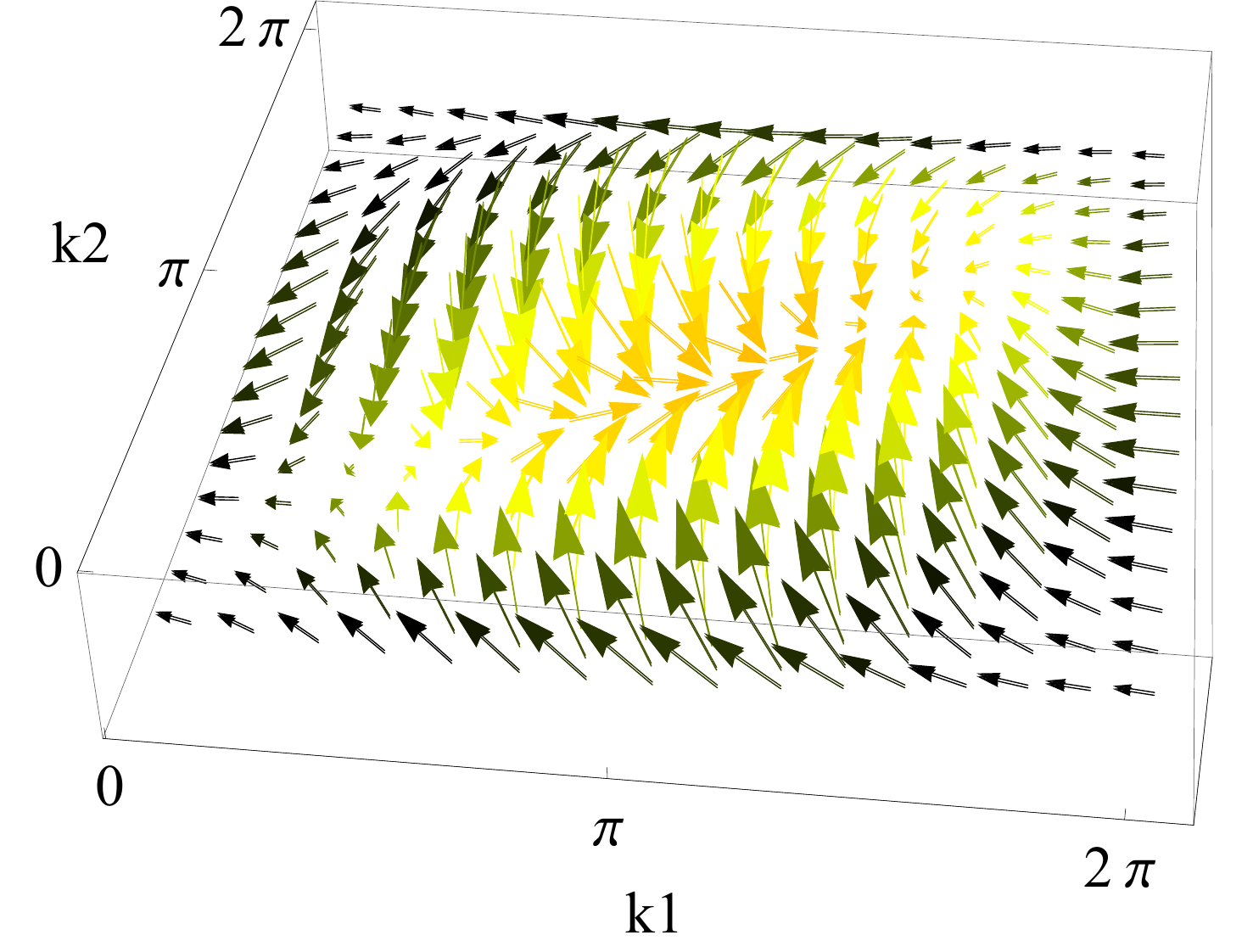} \\ 
\bottomrule 
\end{tabular}
\caption{(Color online) Illustration of our three ideal 2-band FCI models, as well as how they compare to existing 2-band FCI models in the literature. The ideal isotropic FCI condition imposes the constraint that $\hat d$, $\partial_x \hat d$ and $\partial_y \hat d$ (Eq. \ref{dtriad} or equivalently \ref{dtriad2}) must form a mutually orthogonal triad, leading to a more ''fluffed up'' and regular looking $\vec d$-vector texture. By contrast, the textures for the Checkerboard\cite{sun-11prl236803} and Honeycomb\cite{sheng-11ncomm389} models are allowed to contain lines of approximately parallel vectors.   }
\label{fig:table}
\end{table*} 

\section{Examples}
\label{sec:examples}
   
We now turn to demonstrating our ideal FCI engineering procedure through a few examples. The 2-band models with the smallest Berry curvature nonuniformity $\langle(\Delta F)^2\rangle$ are shown to coincide with  simple or well-known models. Models with $N=3$ or more bands can possess arbitrarily low $\langle(\Delta F)^2\rangle$, and we provide explicit constructions of them that possess Berry curvature and $Tr\,g$ that are far more uniform than any hitherto existing model in the literature.

\subsection{$N=2$ band models}

For two bands, there exists only a single elliptic function $\phi_1(z)$ to be tuned. As such, the overall displacement of its singularities is immaterial, and for any given pole structure we are left with a single parameter $\alpha$ denoting the scale of $\phi_1(z)$ relative to the constant $\phi_2(z)=1$ . Despite the highly singular behavior near the poles and restrictions imposed by the ideal isotropic FCI condition, there exist certain $\alpha$'s that yield 2-band models with superior $F_{xy}$ uniformity compared to those in the literature. Such optimal $\alpha$ can be easily found through a one-parameter search. Nevertheless, the choice of two bands precludes a globally-uniform $F_{xy}$, entailing at minimum a single point with $F_{xy} = 0$ in the BZ. This situation will be remedied for $N > 2$ below.

\subsubsection{Model with $\phi_1=5.74\,W$, $C=-2$}

We start with the simplest possible ansatz of $\phi=(\phi_1,1)$ with $\phi_1(z)=\alpha W(z)$, which has a double pole at $z=0$. Numerically, $\langle(\Delta F)^2\rangle$ is minimized when the parameter $\alpha \sim 5.74$. In fact, 
one can show analytically for the square lattice that the optimal ansatz is
\begin{align}
	\phi(z) = \left[ \frac{W(z)}{W(\pi)} ~,~~1 \right]^\top
	\label{Wz}
\end{align}
This corresponds to the Hamiltonian
\begin{equation}
H=\frac1{1+|\phi_1|^2}\left(\begin{matrix}
 & 1 & -\phi_1 \\
 & -\phi_1 & |\phi_1|^2\\
\end{matrix}\right)
\end{equation}
with $\phi_1=W(z)/W(\pi)$. After truncating the opposite-spin ($\sigma_1$ and $\sigma_2$) hoppings to the nearest-neighbor (NN) terms, the Hamiltonian becomes $H=d_0I+\sum_i d_i\sigma_i$, where
\begin{align}
 d_0 &= 0.27 \cos 2k_x\cos 2k_y\notag\\
 d_1 &= \cos k_x -\cos k_y \notag\\
 d_2 &= \cos k_x + \cos k_y \notag \\
 d_3 &= 2\sin k_x \sin k_y
\label{model1} 
\end{align}
This truncated model has a flatness ratio $f\approx 85$, and Berry curvature nonuniformity $\langle (\Delta F)^2 \rangle =\frac{1}{4\pi^2}\int (F_{xy}-\frac{C}{2\pi})^2d^2k = 0.0253$ (for reference the mean Berry curvature is $\langle F\rangle=\frac{2}{2\pi}=0.318$). It has a Dirac point at $k=(0,0)$ of d-wave type, where $\vec d$ has a winding number of $2$ around $k=(0,0)$, as evident in Table. \ref{fig:table}.

It has comparable $\langle (\Delta F)^2 \rangle$ with some $
|C|=1$ models in the literature, such as the Dirac and Honeycomb models with $\langle (\Delta F)^2 \rangle=0.0423$ and $\langle (\Delta F)^2 \rangle=0.0198$ respectively. This is in spite of the more complicated $\vec d$ winding of our Chern number $|C|=2$ model\footnote{Although in princple, the maximal Berry curvature uniformity increases with higher Chern number, as there will be more room to optimize the poles of the ellptic functions.}. 

\subsubsection{Model with $\phi_1=0.85\frac{W'}{W}=0.85\,\partial_z (\log W)$, $C=-2$}

Here we optimized a slightly more complicated ansatz $\phi_1(z)=\alpha\frac{W'(z)}{W(z)}$, where $\alpha$ is optimal at $0.85$. It has order 1 poles at $z=0$ and $z=\pi(1+i)$, and is also proportional to $\zeta(z)+\zeta(z-\pi(1+i))$. 

After an analogous real-space truncation, we obtain $H=d_0I+d\cdot \sigma$, where
\begin{align}
d_0 &= 0.42 \sin^2 k_x\sin^2 k_y,\notag\\
d_1 &= \sin k_x, \notag\\
d_2 &= \sin k_y, \notag\\
d_3 &= \cos k_x \cos k_y .
\label{model2}
\end{align}
This is another type of d-wave model with a higher flatness ratio $f=110$, and comparable $\langle (\Delta F)^2 \rangle = 0.0268$. The Berry curvature uniformity is suspiciously close to that of the previous case ($\langle (\Delta F)^2 \rangle = 0.0253$), and in fact it seems generically true that ans\"atze with the same number of poles, but at different positions, possess almost equal optimal Berry curvature uniformity. 

\subsubsection{Model with $\phi_1=0.7 (\frac{W'}{W})^2$, $C=4$}

Here we provide an example of a less trivial 2-band model with a higher Chern number $C=4$. It uses an ansatz that is the square of that of the previous $C=2$ example. After real-space truncation, we obtain $H=d_0I+d\cdot \sigma$, where
\begin{align}
d_0 &= 0.072 (\cos 4k_x + \cos 4k_y),\notag\\
d_1 &= \sin k_x^2-\sin k_y^2, \notag\\
d_2 &= \sin k_x \sin k_y,\notag\\
d_3 &= -\cos k_x \cos k_y. 
\label{model3}
\end{align}
It has a high flatness ratio $f=120$ and $\langle (\Delta F)^2 \rangle = 0.101$, whose underlying $\vec d$-vector field is superficially similar to a compressed version of that of the previous $C=-2$ model (Table. \ref{fig:table}). 

\subsection{$N=3$ band models}
\label{N3}
Here our ansatz contains two elliptic functions $\phi_1$ and $\phi_2$, and the total Chern number is the sum of the total degree of their poles. With the instantons defined separately on two functions, one can design special configurations where the singular behavior of one of them almost cancels that of the other one. 

\subsubsection{Model with $\phi_1=0.88 \frac{W'}{W}$, $\phi_2=4.3 W$, $C=3$}

This is a model that combines the elliptic functions of two of the 2-band models discussed above. With newly optimized coefficients, it inherits the $F_{xy}$ and $\textrm{Tr} \, g$ uniformity of their predecessors, and in fact fares much better. Overall, the unnormalized Bloch state displays a double pole at $z=0$ and a single pole at $z=\pi(1+i)$, leading to a Chern number of 3. Before real-space truncation, the Berry curvature nonuniformity is $\langle (\Delta F)^2 \rangle =\frac1{4\pi^2}\int (f-\frac{C_1}{2\pi})^2d^2k = 0.00656$, which is a significant improvement from the previous examples. 
\begin{figure}[h]
 \includegraphics[width=0.53\linewidth]{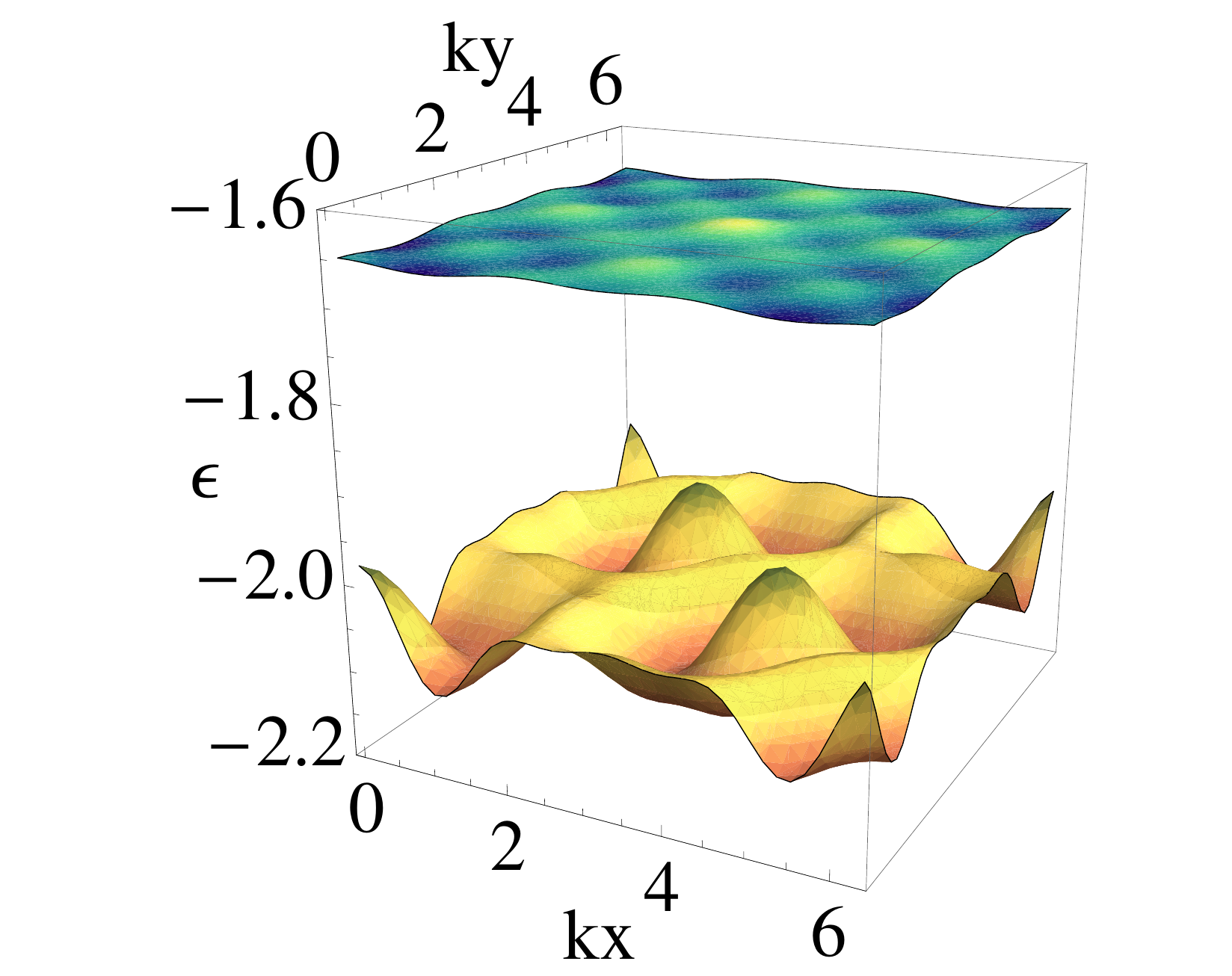}
 \includegraphics[width=0.45\linewidth]{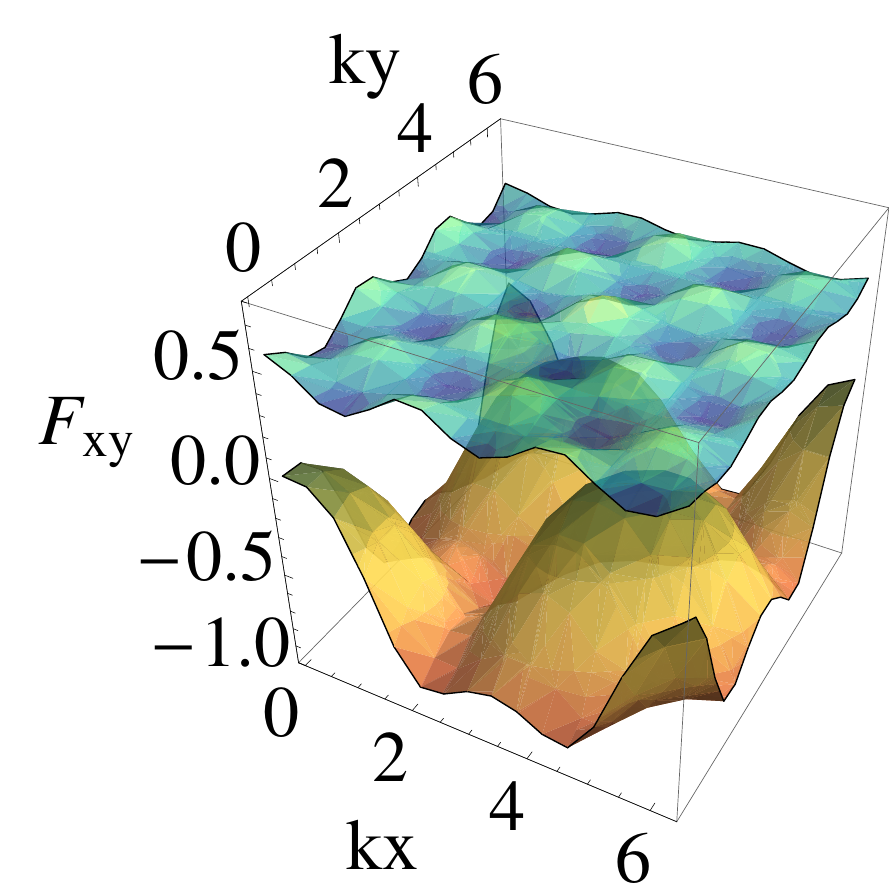}
\caption{(Color Online) Comparison of our truncated $\phi_1=0.88 \frac{W'}{W}$ and $\phi_2=4.3 W$ model (Eqs. \ref{C3_H1} and \ref{C3_H2}) with a well-known optimized FCI model with Chern number $3$ based on pyrochlore slabs (Ref. \onlinecite{trescher2012flat}). In both the Left panel (band dispersion) and Right panel (Berry curvature), our model (bluish-green) is seem to possess much better uniformities than that from Ref. \onlinecite{trescher2012flat} (yellow). Numerically, we have $f\approx 40$ and $\langle (\Delta F)^2\rangle = 1.562\times 10^{-3}$.}
\label{fig:C3compare}
\end{figure}
 A real-space truncation generically breaks the holomorphicity of the Hamiltonian, thereby leading to slightly unequal $\textrm{Tr} \, g \neq F_{xy}$. By truncating to only NN and NNN terms, we arrive at a model with remarkably ideal FCI properties (Fig. \ref{fig:C3compare}), with high flatness ratio $f\approx 40$ and, at the same time, only minimal Berry curvature nonuniformity $\langle (\Delta F)^2\rangle = 1.562\times 10^{-3}$ and $\langle (\Delta \textrm{Tr} \, g)^2\rangle = 1.09\times 10^{-2}$. The Berry curvature uniformity is more than an order of magnitude higher than optimized models in the literature\cite{trescher2012flat}, and is in fact also much higher than that of its non-truncated version. Furthermore, $\textrm{Tr} \, g$ and $F_{xy}$ are still almost equal, with $\langle (\Delta (F-\textrm{Tr} \, g))^2\rangle/\langle F^2\rangle = 1.584\times 10^{-4}/9=1.76\times 10^{-5}$.

\begin{widetext}
\begin{figure*}
	\includegraphics[width=\textwidth]{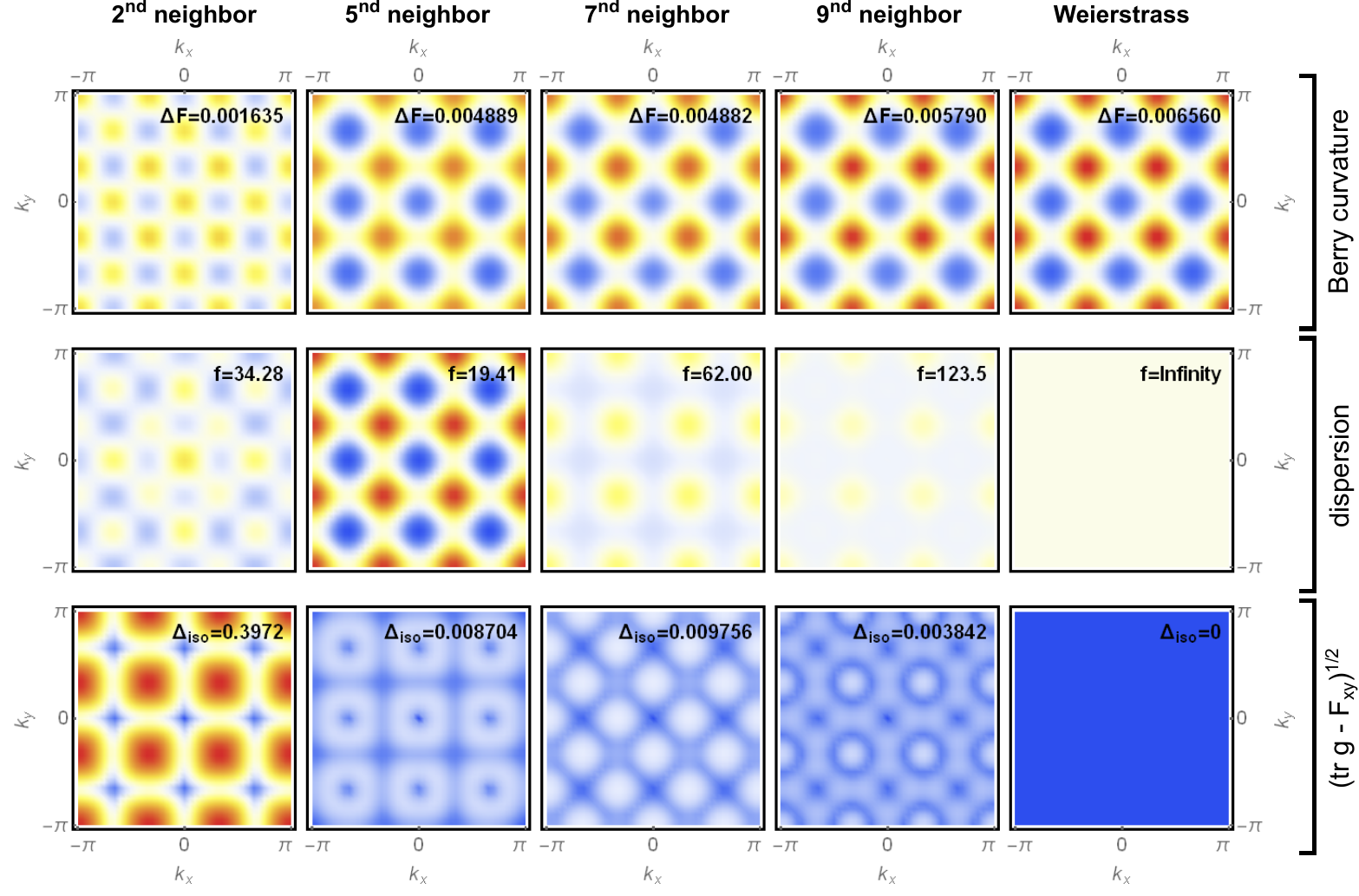}
	\includegraphics[width=10cm]{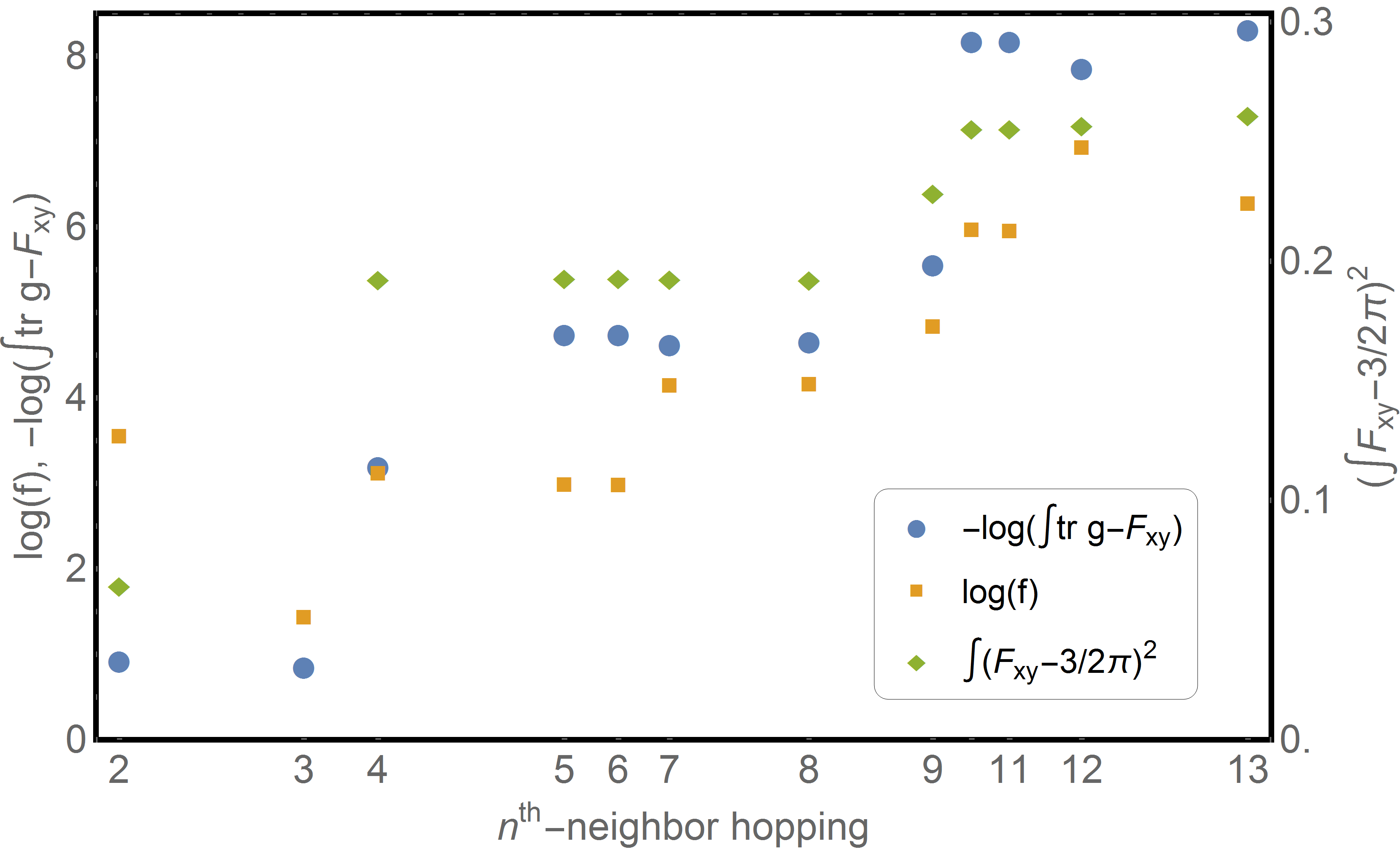}
	\caption{Illustration of the truncation procedure results for the $C=3$ ideal FCI model in section \ref{N3}. The top, middle, and bottom rows depict the Berry curvature, Chern band dispersion, and satisfaction of the ideal droplet condition for our untruncated ideal FCI model (right-most column ``Weierstrass'') as well as real-space truncations to second-, fifth-, seventh- and ninth-neighbor hopping (left and center columns). Inset labels quantify the Berry curvature deviation $\Delta F = (1/4\pi^2) \int d^k (F_{xy} - 3/2\pi)^2$, band flatness ratio, and deviation from the ideal isotropic droplet condition $\Delta_{\rm iso} = \int d^k (Tr~ g - F_{xy})$. These quantities are systematically plotted as a function of real-space truncation length in the bottom plot. Notably, Berry curvature fluctuations are suppressed for shorter-ranged truncations, however at the price of breaking FCI isotropy.} 
	\label{fig:hoppingAndTruncation}
\end{figure*}
\end{widetext}



In real-space, our 3-band ideal FCI model takes the following form. Denote its hopping elements as $H_{ij}(d_x,d_y)$, where $i,j=a,b,c$ represent the atoms within each unit cell, and $d_x\hat x+d_y\hat y$ represent the displacement between the unit cells. We obtain
\begin{equation}
H(1,0)=\left(
\begin{array}{ccc}
 6.65 & 4 i & -6 \\
 4 i & -4.87805 & 7.57576 i \\
 -6 & 7.57576 i & -1.78571
\end{array}
\right)
\label{C3_H1}
\end{equation}
and
\begin{eqnarray}
&&H(1,1)=\nonumber \\
&&\left(
\begin{array}{ccc}
 -1.10833 & -3.125(1+i) & -3.5 i \\
 3.125(1-i) & -2.94118 & 1.66667(1+ i) \\
 3.5 i & -1.66667(1-i) & 4
\end{array}
\right).\nonumber\\
\label{C3_H2}
\end{eqnarray}
Hoppings to the other NN and NNN unit cells are related by the following symmetries: Upon a spatial rotation of $\pi/2$ clockwise (i.e. $(1,1)\rightarrow (-1,1)$), $H_{aa},H_{bb},H_{cc}$ remained unchanged, $H_{ab},H_{bc},H_{ba},H_{cb}$ are multiplied by $-i$ and $H_{ac},H_{ca}$ are multiplied by $(-i)^2=-1$. No onsite hoppings elements $H(0,0)$ are required to maintain the band flatness.

\subsection{$N>3$ band models with higher Chern number}


Progressively higher Chern numbers introduce additional variational degrees of freedom to the elliptic-functions ansatz, and allow for substantially better uniformity of $F_{xy}$. For $C=4$, an ansatz with $4$ bands
\begin{align}
	\phi(z) = \left( 33.7 W^2 - 0.18, ~ -9W', ~  9W, ~ 1 \right)^\top
\end{align}
gives significantly reduced fluctuations of $F_{xy}$, with $\langle (\Delta F)^2 \rangle = 10^{-3}$. For $C=5$, we arrive at the optimum 
\begin{align}
	\phi(z) = \left( 40.5 W' W, ~ 51.7 W^2 - 0.22, ~ 10.5 W', ~ 10.2 W \right)^\top
\end{align}
with extremely low $\langle (\Delta F)^2 \rangle = 10^{-4}$. We note that the hopping range for the corresponding Hamiltonian increases with increasing Chern number. While the magnitude of hopping matrix elements decays exponentially with increasing hopping range on the lattice, the Berry curvature and Fubini-Study metric remain sensitive to longer-ranged perturbations. This is exemplified in Fig.~\ref{fig:hoppingAndTruncation} for the case of the $C=3$ model introduced above: while the Chern number remains robust already when truncating to second-nearest neighbor hopping, $F_{xy}$ and $\textrm{Tr}\,g$ show significant sensitivity to the hopping range until approaching the expected values for the ideal long-range model, that is, for truncations to 6th neighbor hopping or higher.

\section{Conclusion}
We have introduced a systematic approach for the explicit construction of ``ideal'' fractional Chern insulator bands on which quantum Hall pseudopotentials faithfully take the form of local (and thus more realistic) density-density interactions. This construction is inspired by interchanging the roles of position and momentum when transcending from the real-space magnetic field of FQH systems to the Berry curvature of FCI systems. A central consequence of this position-momentum dual picture is the  {ideal isotropic FCI condition} $\textrm{Tr}\,g = F_{xy}$ which is generically satisfied by Bloch bands that are meromorphic (dependent on $k_x+ik_y$) before normalization. Being confined to the BZ torus, they can be described by simple linear combinations of doubly periodic complex (elliptic) functions.

As such, the parameter space of possible candidate Hamiltonians is greatly reduced to not more than a few independent parameters. This allows important constraints like maximally uniform band dispersion, Berry curvature $F_{xy}$ and $\textrm{Tr}\,g$ to be conveniently met without extensive numerical search. The constraint of maximal band flatness quenches unwanted interference of FQH physics from single-body energetics, and can be systematically optimized through the imaginary gap. This is assisted by the meromorphicity of the Bloch bands, which already minimizes the mean-square distance of the long-range hopping terms that should be truncated for a realistic model. The constraint of maximally uniform Berry curvature, and hence $\textrm{Tr}\,g$, ensures that magnetic translation symmetry is maximally retained in our FCI models, with the density algebra also agreeing with that of FQH systems up to third order. 

Finally, we demonstrate our construction for a series of examples. As one of the highlights, we find a 3-band, $C=3$ model that simultaneously possesses band dispersion and berry curvature uniformities far superior to models in the literature of comparable locality. With almost trivial modifications, our approach may be modified to account for various specific requirements, such as to give higher priority to either band dispersion of Berry curvature uniformity, or adherence to specific symmetries. Altogether, we believe that our construction establishes the most convenient and powerful approach to perfrom band structure engineering for ideal fractional Chern insulators.


\begin{acknowledgements}
We thank Bo Yang, Chaoming Jian and Emil Bergholtz for helpful discussions. This work was supported by the German Science Foundation through DFG-SFB 1170 (project B04) and the European Research Council through ERC-StG-Thomale-336012-TOPOLECTRICS.

\end{acknowledgements}

\newpage
\appendix

\section{Constraints on poles and zeros of elliptic functions}
\label{sect:elliptic}
Consider a meromorphic $\phi(z)$ which is doubly periodic with periods $(1,\tau)$, i.e. that $\phi(z)=\phi(z+1)$ and $\phi(z)=\phi(z+\tau)$. The unit cell boundary $\Gamma=\partial \Gamma$ consists of line segments $[0,1],[1,1+\tau],[1+\tau,\tau]$ and $[\tau,1]$. Since the opposite sides of $\Gamma$ give equal but opposite contributions, the residue theorem yields
\begin{equation}
\oint_\Gamma \phi(z)dz = 2\pi i \sum_j Res(\phi(z_j))=0
\end{equation} 
i.e. $\phi(z)$ has residues summing to zero. Employing the same trick to $\partial_z\log \phi(z)$, we obtain 
\begin{equation}
\oint_\Gamma \partial_z\log\phi(z)dz =\sum_i Z_i - \sum_k R_k =0
\end{equation}
where the $Z_i$'s and $R_k$'s are respectively the orders of the zeros and poles of $\phi(z)$. Hence the number of zeros and poles of $\phi(z)$ are equal, with multiplicities counted.

Combining these two observations, we see that it is impossible for a doubly periodic $\phi(z)$ to have only one pole. Hence the FCI model constructed from it must necessarily possess a Chern number $C>1$.

\section{Complex analytic properties of flat-band ideal FCI Hamiltonians}
\label{sect:holoham}

Recall from Eq. \ref{Hmn} that our ideal FCI Hamiltonians take the form
\begin{eqnarray}
H_{mn}(k_x,k_y)&=&\delta_{mn}-\varphi_m(k_x,k_y)\varphi_n^*(k_x,k_y)\notag\\
&=&\delta_{mn}-\frac{\phi_m(k_x,k_y)\phi_n^*(k_x,k_y)}{|\phi(k_x,k_y)|^2}\notag\\
&=&\delta_{mn}-\frac{\phi_m(k_x+ik_y)[\phi_n(k_x+ik_y)]^*}{|\phi(k_x+ik_y)|^2}\notag\\
\label{Hmnapp}
\end{eqnarray}
On the last line, we have used the fact that $\phi_m,\phi_n$ depend holomorphically on $k_x$ and $k_y$. The Hamiltonian can be further analytically continued into the complex plane in each of these variables separably. Explicitly, we can either perform the replacement $H(k_x,k_y)\rightarrow H(k_x+ig_x,k_y)$ or $H(k_x,k_y)\rightarrow H(k_x,k_y+ig_y)$, where $g_x,g_y$ are real. Let $h(k_x,k_y)$ be a matrix element of the original Hamiltonian, not necessarily of holomorphic type. Write $\tilde h(z,k_y)$ and $\tilde h(k_x,z)$ be their (in general different) analytically continued versions into the complex $k_x$ and $k_y$ planes respectively. A few examples:
\begin{itemize} 
\item If $h(k_x,k_y)=\sin k_x+ i\sin k_y$, i.e. a p-wave term, the (different) analytic continuations in the $k_x$ and $k_y$ planes are given by $\tilde h(z,k_y)=\sin z+i\sin k_y$ and $\tilde h(k_x,w)=\sin k_x+i\sin z$ respectively.

\item If $H(k_x,k_y)$ is the Weierstrass model (Eq. \ref{Wz}) with $\phi_1(z)=\alpha W(z)$, $\phi_2(z)=1$, the off-diagonal matrix element is given by $h(k_x,k_y)=\frac{\alpha W(k_x+ik_y)}{1+\alpha^2|W(k_x+ik_y)|^2}$ where $W(z)$ is the Weierstrass function containing a double pole around the origin. $W(z)$ possess the property that $[W(z)]^*=W(z^* )$, so $|W(k_x+ i k_y)|^2= W(k_x+ik_y)W(k_x-ik_y)$ for $k_x,k_y\in \mathbb{R}$. Setting $\alpha=1$ for brevity, the normalized Bloch eigenfunction $\varphi_1,\varphi_2$ are, \textit{before} analytic continuation,
\begin{equation} \varphi_1(k_x,k_y)=\frac{W(k_x+i k_y)}{\sqrt{1+W(k_x+ik_y)W(k_x-ik_y)}} \label{varphi1}\end{equation}  
\begin{equation} \varphi_2(k_x,k_y)=\frac{1}{\sqrt{1+W(k_x+ik_y)W(k_x-ik_y)}} \label{varphi2}
\end{equation}  
After analytic continuation into say, the $k_x$-direction, these eigenfunctions define, via Eq. \ref{Hmnapp}, the Hamiltonian matrix element 
\begin{equation}
\tilde h(z,k_y)=\frac{W(z+ik_y)}{1+W(z+ik_y)W(z-ik_y)}
\label{hkxz}
\end{equation}
The real space decay rate $g=g_x=g_y$ of the Hamiltonian matrix elements are equal due to $x\leftrightarrow y$ symmetry, and is given by the imaginary gap, i.e. the imaginary part of the singularity closest to the real axis, where the denominator of Eq. \ref{varphi1} or \ref{varphi2} is zero. Focusing on the $k_x$ direction without loss of generality, $g_x=g_x(k_y)$ is the root with smallest magnitude satisfying
\begin{equation}
1+W((k_x+ig_x)+ik_y)W((k_x+ig_x)-ik_y)=0\notag\\
\label{eq1}
\end{equation}
for some $k_y$. Here it is useful to think of $W$ as a holomorphic function of $k_x+ig_x$, $k_x,g_x\in \mathbb{R}$, with $k_y$ regarded as a real parameter. The roots of Eq. \ref{eq1} correspond to the three singularities of Eq. \ref{hkxz}, which are connected by a branch cut as shown in Fig. \ref{weierstrass}. The imaginary gap is the minimal distance of these singularities from the real axis, i.e. the minimal $g_x$.
\begin{figure}[H]
\includegraphics[scale=0.41]{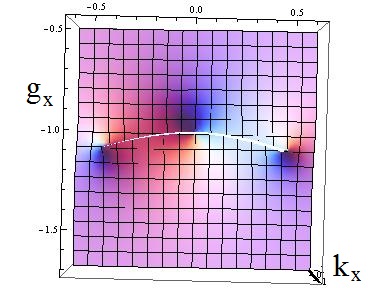}
\includegraphics[scale=0.4]{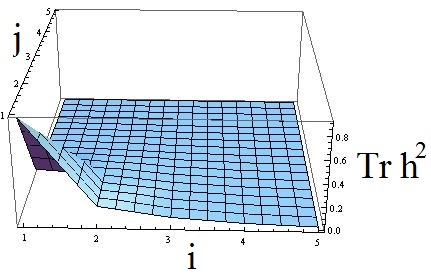}
\caption{(Color online) Left) The branch cuts and 3 singularities of Eq. \ref{hkxz}. There is also an additional singularity when $g_x=\pm k_y$ since $W(x)$ diverges at $x=0$.  Right) $\sqrt{\textrm{Tr}{h^\dagger_{ij}h_{ij}}}$ for $|i-j|$ =$0$ to $4$ . Terms beyond NNs and NNNs are more than 1 order of magnitude smaller than the onsite term. 
\label{weierstrass}
}\end{figure}

\end{itemize}
\subsection{Locality of the Hamiltonian}

The exponential decay of the real space hoppings in our models is fortunately fast enough for hoppings beyond the first few to be safely truncated without introducing significant bandstructure deviations. However, an upper limit exist for the decay rate of such holomorphic models due to the compactness of the torus BZ. Indeed, the smallest distance of a singularity from the real axis cannot, after averaging over $k_y$, be further than $\pi$, which is half a period on the BZ. 

For more general Bloch eigenfunctions that do not depend holomorphically on $k_x + ik_y$, the requirement for double periodicity is lifted. Hence it is possible that the poles are much further away than $\pi$ from the real axis, when $k_x$ and $k_y$ are considered separate holomorphic functions. However, this does not appreciably improve the band flatness in any practical sense, since a NNN truncation of a model with $g=\pi$, for instance, already has a superb approximate flatness ratio of $e^{2\pi}\approx 500$. 

\subsection{Further properties of Hamiltonians with holomorphic dependence $k_x+ik_y$}
 A general Hamiltonian with elements of the form $h(k_x+ ik_y)$ must be doubly periodic, and can hence be described in terms of the Weierstrass function $W(k_x+ik_y+a)$ and its derivative $W'(k_x+ik_y+a)$, where $a$ is a constant complex quantity that sets the position of the poles. The following nice property that 
\begin{equation} [W(k_x+ik_y+a)]^*=W(k_x-ik_y+a^*) \end{equation}
holds for $W$ and also analogously for $W'$, so that a function $h$ which depend on $W$,$W'$ and their complex conjugates can in general be written as combinations of Weierstrass functions of both $k_x+ik_y$ and $k_x-i k_y$. Such $h$ enjoy the following symmetry properties:
\begin{itemize}
\item \textbf{Invariance of decay rate under translation of poles:} Under a translation by a complex $a$, 
\begin{equation} k_x+ ik_y\rightarrow k_x + ik_y+a = (k_x+\text{Re}\, a)+ i (k_y + \text{Im}\ a)\end{equation}
\begin{equation} k_x- ik_y\rightarrow k_x - ik_y+a^* = (k_x+\text{Re}\ a)- i (k_y + \text{Im}\ a)\end{equation}
This is merely a redefinition of $k_x$ and $k_y$ whose effect can be easily absorbed by the summation $\sum_{k_xk_y}$. Hence the inverse decay lengths $g_x(k_y)$ and $g_y(k_x)$ will not be affected by a uniform translation of $a$ in all Weierstrass functions involved.
\item \textbf{Conditions for $x\leftrightarrow y$ symmetry} 
There exist further restrictions on the functional form of the Hamiltonian if the system is symmetric under an interchange of $x$ and $y$ (or $k_x$ and $k_y$), which implies the equivalence of imaginary gaps $g_x$ and $g_y$. From the relations $W(iz)=-W(z)$ and $W'(iz)=iW'(z)$, we obtain
\begin{equation} W(k_x\pm i k_y+a)=-W(k_y\mp i k_x\mp ia)\end{equation}
\begin{equation} W'(k_x\pm k_y +a)=\mp i W'(k_y\mp ik_x \mp ia) \end{equation}
for all real $k_x,k_y$ and complex $a$. Hence if $a\neq 0$, then a Hamiltonian must also contain other $ia$,$-a$, $-ia$, i.e. be C4 symmetric in $a$ for it to be invariant under $x\leftrightarrow y$ symmetry. Furthermore, common factors of the Hamiltonian must also contain terms $|W(k_x+ik_y+a)|^2$ or $|W'(k_x+ik_y+a)|^4$ and their C4 related terms for $x\leftrightarrow y$ symmetry to hold. For instance, $(1+W(k_x+ik_y)W(k_x-ik_y))^{-1}$ is a valid possibility for a Hamiltonian matrix element but not $(1+W(k_x+ik_y)W'(k_x-ik_y))^{-1}$.

\end{itemize}

\section{The K\"ahler potential of $F_{xy}, \textrm{Tr}\, g$ and its relation to ideal isotropic FCIs}
\label{sect:berry}

With the holomorphic ansatz $\phi(z)$
, the ideal isotropic FCI condition $F_{xy}=\textrm{Tr}\, g$  holds, established by Eqs. \ref{Fg}. Let us express it in an alternative form which bears direct relevance to ideal isotropic FCIs. Since $\partial_{\bar z}\phi_i=0$, the gauge connection $A_j=-i\varphi^\dagger\cdot \partial_j \varphi$ satisfies the Coulomb gauge condition $\partial\cdot A=0$, as is most easily seen in the $z,\bar z$ coordinates. 
This allows us to write $A_x=\partial_y \Phi$, $A_y=-\partial_x \Phi$, where 
\begin{equation} 
\Phi = \frac{1}{2}\log \left(\sum_i |\phi_i(z)|^2\right)
\end{equation}
is the K\"ahler potential satisfying $F_{xy}=\nabla^2 \Phi=4\partial_{\bar z}\partial_z \Phi$. This holds true whether we have imposed $\phi_N=1$ or not. Since $\nabla^2\log|\phi_i(z)|^2=0$ for each component $i$, the K\"ahler potential and thus $F_{xy}$ arises solely from the ``interactions'' between the different components brought about by the nonlinearity of the logarithm.  

For our purpose, we strive to find $\Phi(z)$ whose Laplacian $F_{xy}$ is as uniform as possible. From the study of ideal isotropic FCIs\cite{claassen2015}, we already know of an ansatz with perfectly uniform Laplacian, i.e:
\begin{equation} 
\sum_i |\phi_i|^2 = \left(e^{-\frac{|z+\bar z|^2}{8\pi}}|\theta_j (iz/2;e^{-\pi})|^2\right)^C
\label{Laplacian2}
\end{equation}
where $C$ is the Chern number and $\theta_j$, $j=2,3$ are the Elliptic Theta functions. Similar expressions exist for $j=1,4$.

In general, $\sum_i |\phi_i|^2$ has $C$ poles for Chern number $C$, each of which can be placed arbitrarily by generalizing the ansatz above. The optimization problem for uniform $F_{xy}$ can thus be recast into finding combinations of $\zeta(z-b_j), W(z-b'_j)$ that optimally reproduce the ideal K\"ahler potential above. It can be determined by inspection that a Laurent series expansion around each pole of the K\"ahler potential can be matched only by an infinite superposition of Zeta functions and Weierstrass elliptic functions with the same pole. Therefore, the simplest strategy places a single $C$th-order pole in the K\"ahler potential such as to employ all available elliptic functions to approximate the behavior around this pole.

\section{Parametrization and geometric properties of 3-component models}
\label{sect:Trg}

Here we provide some details on the parametrization of a 3-component Hamiltonian that is analogous to the Bloch sphere parametrization of a 2-component model.

An eigenstate $w$ of a 3-band models lives in the state space $\mathbb{CP}^{2}$, which we parametrize as 
\begin{equation}
w=(ae^{i(\mu+\lambda_1)},be^{i(\mu+\lambda_2)},ce^{i\mu})
\end{equation}
with real parameters $a,b,c,\lambda_1,\lambda_2$ such that $a^2+b^2+c^2=1$, i.e $(a,b,c)\in S^2$ and $(\lambda_1,\lambda_2)\in T^2$. Of course, $\mathbb{CP}^2$ is not a product space: $\mathbb{CP}^{2}\neq S^2\times T^2$, just like $\mathbb{CP}\neq S^1\times S^1$. Indeed, there exist certain points on $\mathbb{CP}^{2}$, namely those where at least one of $a,b,c$ is zero, where all points in the $T^2$ fiber collapses as a single point on $S^2$. This is in analogy to the north and south poles of $\mathbb{CP}\sim S^2$. 

The Fubini-Study metric can be shown to take the form
\begin{equation}
ds^2 = a^2 (1-a^2) d\lambda_1^2 + b^2 (1-b^2)d\lambda_2^2 -2a^2b^2 d\lambda_1d\lambda_2 + d\Omega^2
\end{equation}
where $d\Omega^2=d\theta^2 + \sin^2\theta d\phi^2$ is the standard metric on the sphere $S^2$. Note that the overall phase $\mu$ drops out completely.

The Berry curvature $F_{xy}$ can be expressed as the winding of a certain higher-dimension ``d-vector'' on the 7-sphere\cite{lee2015arbitrary} $S^7$ upon switching to the basis spanned by the Lie Algebra generators of $SU(3)$.  

\bibliographystyle{prsty}
\bibliography{paper,TI,thesis,references,pseudopotentials}

\end{document}